\documentclass[onecolumn]{aastex62}
\usepackage{amsmath}     
\usepackage{wasysym}           
\usepackage{graphicx}
\usepackage{amssymb}
\usepackage{epstopdf}
\usepackage{mathrsfs}
\usepackage{anyfontsize}
\usepackage{natbib}
\usepackage{color}
\usepackage{lipsum}
\usepackage{diagbox}

\shorttitle{Energy-conserving, Relativistic Corrections to Strong Shock Propagation}
\shortauthors{Coughlin, E.R.}
\begin{document}
\title{Energy-conserving, Relativistic Corrections to Strong Shock Propagation}
\author[0000-0003-3765-6401]{Eric R.~Coughlin}
\altaffiliation{Einstein Fellow}
\affiliation{Columbia Astrophysics Laboratory, New York, NY 80980}

\email{eric.r.coughlin@gmail.com}

\begin{abstract}
Astrophysical explosions are accompanied by the propagation of a shock wave through an ambient medium. Depending on the mass and energy involved in the explosion, the shock velocity $V$ can be non-relativistic ($V \ll c$, where $c$ is the speed of light), ultra-relativistic ($V \simeq c$), or moderately relativistic ($V \sim few\times 0.1c$). While self-similar, energy-conserving solutions to the fluid equations that describe the shock propagation are known in the non-relativistic (the Sedov-Taylor blastwave) and ultra-relativistic (the Blandford-McKee blastwave) regimes, the finite speed of light violates scale invariance and self-similarity when the flow is only mildly relativistic. By treating relativistic terms as perturbations to the fluid equations, here we derive the $\mathcal{O}(V^2/c^2)$, energy-conserving corrections to the non-relativistic, Sedov-Taylor solution for the propagation of a strong shock. We show that relativistic terms modify the post-shock fluid velocity, density, pressure, and the shock speed itself, the latter being constrained by global energy conservation. We derive these corrections for a range of post-shock adiabatic indices $\gamma$ {(which we set as a fixed number for the post-shock gas)} and ambient power-law indices $n$, where the density of the ambient medium $\rho_{\rm a}$ into which the shock advances declines with spherical radius $r$ as $\rho_{\rm a} \propto r^{-n}$. For Sedov-Taylor blastwaves that terminate in a contact discontinuity with diverging density, we find that there is no relativistic correction to the Sedov-Taylor solution that simultaneously satisfies the fluid equations and conserves energy. These solutions have implications for relativistic supernovae, the transition from ultra- to sub-relativistic velocities in gamma-ray bursts, and other high-energy phenomena.
\end{abstract}


\section{Introduction}
\label{sec:intro}
A core-collapse supernova is, in the now-classic picture, initiated by the ``bounce'' of the overpressured, protoneutron star that forms from the collapse of the iron core of a massive star \citep{colgate66}. From the protoneutron star bounce is launched a shock wave, which propagates through and unbinds the stellar envelope to yield the supernova. If it is sufficiently energetic, the shock promptly plows through the star following the bounce and liberates the gas; alternatively, if it is not energetic enough to overcome the ram pressure of the infalling material and the dissociation of heavy nuclei in the core of the star \citep{arnett82}, the shock ``stalls'' at small radii but can be revived by some means (e.g., neutrino heating, convective instabilities behind the shock, a dynamic instability of the standing shock, or the magnetorotational amplification of magnetic fields; respectively, e.g., \citealt{bethe85, burrows95, blondin03, mosta15}). If the shock fails to be revived, with sufficient angular momentum an accretion disc forms around the natal black hole that can, through the combination of bipolar outflows and liberated accretion energy, unbind the remaining stellar envelope in the collapsar picture of a long gamma-ray burst \citep{woosley93,macfadyen99, woosley06}. Finally, even in the absence of sufficient angular momentum, a failed supernova generates a secondary, weak shock in the outer layers of the star from the mass lost to neutrinos during the de-leptonization of the core \citep{nadezhin80, lovegrove13, piro13, coughlin18a, fernandez18, coughlin18b}. 

All of these explosion scenarios involve the formation and expansion of a shock wave into its surroundings, and this shock leaves in its wake a ``sea'' of post-shock fluid (as, of course, do explosion scenarios not initiated by the collapse of a massive star, such as compact object mergers; e.g., \citealt{li98, levinson02, nakar11, abbott17c}). 
One of the most useful techniques for describing the spatial and temporal evolution of the post-shock gas and of the shock itself is self-similarity. This mathematical technique exploits the scale invariance of the fluid equations and, in the absence of any temporal or spatial scales of the ambient medium, the necessary scale invariance of the solutions to those equations (e.g., \citealt{ostriker88}). 

Among the best-known examples of a self-similar solution to the fluid equations is the Sedov-Taylor (ST) blastwave (\citealt{sedov59, taylor50}). The ST blastwave describes the propagation of an energy-conserving, strong (Mach number much greater than one) shock into an ambient medium that possesses a power-law density profile. The conservation of energy implies that there is a unique shock speed $V$ that can be directly related to the initial energy of the explosion, the impulsive injection of which initiated the explosion in the first place. The ST solution is also non-relativistic, in that the shock speed is assumed to be much less than the speed of light and the energy is Newtonian, and hence terms of order $V^2/c^2$ that enter into the relativistic fluid equations are ignored ($c$ being the speed of light). The ST blastwave can be used to describe terrestrial explosions and can also constrain the age of supernova remnants \citep{chevalier76}. 

In the other, extreme limit of an ultra-relativistic explosion -- where the shock velocity is nearly equal to the speed of light -- \citet{blandford76} derived a distinct, energy-conserving, self-similar solution to the relativistic fluid equations. The Blandford-McKee (BMK) solution is the ultra-relativistic analog of the Sedov-Taylor blastwave, in that the conservation of energy implies that there is a unique shock Lorentz factor $\Gamma = (1-V^2/c^2)^{-1/2}$ that is relatable to the explosion energy. The BMK blastwave is ultra-relativistic in the sense that the solution only accounts for terms in the fluid equations to order $\mathcal{O}(1/\Gamma^2)$. While there are at present no (known) terrestrial applications of this solution, gamma-ray bursts should exhibit some phase of shock propagation appropriate to ultra-relativistic speeds, and simulations have found evidence of the transition to a Blandford-McKee-type phase of explosion \citep{kobayashi00, duffell13, xie18}.

In this paper we are interested in the behavior of an energy-conserving explosion that is between the limits of Newtonian and ultra-relativistic. The specific question we ask is: when the flow is mildly relativistic, such that the shock speed is $V \sim few\times0.1 c$, how do relativistic effects modify the Sedov-Taylor solution? There are particularly energetic supernovae and lower-luminosity gamma-ray bursts that can produce marginally relativistic shock speeds \citep{soderberg06, drout11, corsi14, corsi16, whitesides17}. For these modestly-relativistic flows, how do relativistic corrections modify the shock velocity scaling predicted from the ST blastwave, and how are the post-shock velocity, density, and pressure profiles altered from the self-similar solutions provided by the ST solution? 

In this paper, we answer these questions by considering special relativistic terms as perturbations to the non-relativistic fluid equations and, correspondingly, the solutions to those equations. In Section \ref{sec:general}, we first give some basic considerations of the problem, and we derive order of magnitude estimates for the corrections to the shock velocity that are induced by relativistic motion. In Section \ref{sec:fluid} we present a rigorous perturbation analysis of the shock jump conditions and the fluid equations to leading relativistic order, and from those equations we derive relativistic, non-self-similar corrections to the post-shock velocity, density, and pressure that result from the velocity scale established by the finite speed of light. We also demonstrate that there is a unique, relativistic correction to the shock velocity that results from the requirement that the energy -- which includes relativistic terms -- be exactly conserved behind the shock front.  We summarize and conclude in Section \ref{sec:summary}.

\section{General Considerations and Order of Magnitude Estimates}
\label{sec:general}
We characterize an explosion by the ejection of an amount of mass $M_{\rm ej}$ with a corresponding energy $E_{\rm ej}$, which can be combined to yield a characteristic velocity $V_{\rm ej} = \sqrt{E_{\rm ej}/M_{\rm ej}}$. As this material encounters an ambient medium medium, a forward shock is generated that initially propagates at the same characteristic speed $V_{\rm ej}$. Once the shock entrains sufficient inertia from its surroundings that the initial mass of the explosion is forgotten, the forward shock propagation settles into a self-similar state such that the boundary conditions at the shock (namely the jump conditions; see Section \ref{sec:jump}) govern the entire post-shock evolution of the flow, while the characteristic ambient density $\rho_{\rm a}$ and length scale $r_{\rm a}$ and the shock energy $E_{\rm ej}$ dictate the propagation of the shock itself. If the density of the ambient medium falls off as a power-law with radial power-law index $n$, then, during this self-similar phase, energy conservation in the non-relativistic limit implies that the shock velocity $V$ is related to the shock position $R$ via

\begin{equation}
V \simeq \sqrt{\frac{E_{\rm ej}}{\rho_{\rm a}r_{\rm a}^3}}\left(\frac{R}{r_{\rm a}}\right)^{\frac{n-3}{2}} \simeq V_{\rm ej}\left(\frac{R}{r_{\rm a}}\right)^{\frac{n-3}{2}}. \label{Vsh}
\end{equation}
The last line in this expression follows from the fact that the self-similar state is only reached once the mass entrained from the ambient medium is comparable to the initial mass of the explosion, i.e., $\rho_{\rm a}r_{\rm a}^3 \simeq M_{\rm ej}$. This expression can be rearranged and integrated to solve for the shock position as a function of time, which yields

\begin{equation}
R \propto t^{\frac{2}{5-n}}. \label{RST}
\end{equation}
This temporal scaling of the shock position is the well-known, Newtonian-energy-conserving result derived independently by \citet{sedov59} and \citet{taylor50}

We see from Equation \eqref{Vsh} that when the mass involved in the explosion is small or the energy imparted to that mass is large, both of which are possible in astrophysical contexts, the characteristic ejecta velocity can exceed the speed of light. In this limit the Newtonian approach breaks down, and instead of being constrained by a characteristic velocity, the shock can be parameterized by its Lorentz factor $\Gamma = (1-V^2/c^2)^{-1/2}$. In the ultra-relativistic limit, the post-shock inertia is dominated by the internal energy of the gas, which scales as $\rho_{\rm a}\Gamma^2$, and mass conservation implies that the post-shock gas is swept into a thin shell of width $\sim R/\Gamma^2$ \citep{blandford76}. To leading order in the shock Lorentz factor, relativistic energy conservation then dictates that the Lorentz factor of the shock satisfies

\begin{equation}
\Gamma \simeq \sqrt{\frac{E_{\rm ej}}{\rho_{\rm a}r_{\rm a}^3c^2}}\left(\frac{R}{r_{\rm a}}\right)^{(n-3)/2} \simeq \frac{V_{\rm ej}}{c}\left(\frac{R}{r_{\rm a}}\right)^{(n-3)/2}. \label{Gammash}
\end{equation}
In this expression $R \simeq c \,t$, and it therefore follows that the Lorentz factor of the relativistic, self-similar shock evolves temporally as

\begin{equation}
\Gamma \simeq \frac{V_{\rm ej}}{c}\left(\frac{ct}{r_{\rm a}}\right)^{\left(n-3\right)/2}. \label{Gammaoft}
\end{equation}

In between these two limits -- Sedov-Taylor when $\sqrt{E_{\rm ej}/M_{\rm ej}} \ll c$ and Blandford-McKee when $\sqrt{E_{\rm ej}/M_{\rm ej}} \gg c$ -- how does the flow behave? In addition to the examples of hyperenergetic supernovae considered in Section \ref{sec:intro}, it can also be seen from Equation \eqref{Gammaoft} that -- even if the blastwave is extremely energetic and begins in the Blandford-McKee regime -- the deceleration of the shock Lorentz factor (assuming $n < 3$, above which the Sedov-Taylor solution does not exist; we will always adopt $n < 3$ in this paper, and return to the case of $n > 3$ briefly in the conclusions) implies that the ultra-relativistic approximation only holds for a finite time $\Delta t_{\rm rel}$.  Setting $\Gamma = 1$ in Equation \eqref{Gammaoft} and rearranging shows that this timescale is

\begin{equation}
\Delta t_{\rm rel} \simeq \frac{r_{\rm a}}{c}V_{\rm ej}^{\frac{2}{3-n}} \simeq \frac{r_{\rm a}}{c}\left(\frac{E_{\rm ej}}{M_{\rm ej} c^2}\right)^{\frac{1}{3-n}}.
\end{equation}
For fiducial values of $E_{\rm ej} \simeq 10^{52}$ erg, $M_{\rm ej} \simeq 0.1 M_{\odot}$, $n =0$, and setting $r_{\rm a} \simeq N \times R_{\odot}$, we find $\Delta t_{\rm rel} \simeq 1\times N$ sec. Therefore, even for large values of $r_{\rm a}$ (or correspondingly small values of the ambient density), the ultra-relativistic phase of self-similar shock propagation can be very short lived. Relativistic terms that modify the solution from the non-relativistic, Sedov-Taylor phase -- to which the flow eventually asymptotes -- will then be present during the transition from ultra- to non-relativistic shock expansion.

At the order of magnitude level, the importance of relativistic corrections to the Sedov-Taylor blastwave can be understood by noting that the leading-order, relativistic modifications to the fluid equations appear as $\propto \mathcal{O}(v^2/c^2)$, where $v$ is the three-velocity of the fluid; we derive these corrections explicitly in Section \ref{sec:fluid}, but such a scaling is reasonable from the observation that the four-velocity, which differs from the three velocity by a factor of $\Gamma \simeq 1+\mathcal{O}(v^2/c^2)$, transforms in a covariant (i.e., tensor-like) sense and therefore enters manifestly into the relativistic fluid equations. There will therefore be corrections of this same order to the jump conditions at the shock front, and hence the lowest-order, relativistic corrections to the post-shock velocity, density, and pressure will be of the order $V^2/c^2$. There will also be modifications to the conserved energy that enter as $v^2/c^2$ (again, we show this explicitly below, but this feature follows naturally from the covariant nature of the four-velocity over the three-velocity). Therefore, in order to satisfy energy conservation, there must also be relativistic corrections of the order $V^2/c^2$ that modify the propagation of the shock itself. We thus expect that during the marginally-relativistic phase when $V_{\rm ej}/c \lesssim 1$, the shock speed will be characterized by

\begin{equation}
V^2 \simeq V_{\rm ej}^2\left(\frac{R}{r_{\rm a}}\right)^{n-3}\left(1+\sigma\frac{V^2}{c^2}\right), \label{Vapp}
\end{equation}
where $\sigma$ is an unknown but otherwise pure number. Since the shock is assumed to be only mildly relativistic, it follows that this relativistic term can be approximated as

\begin{equation}
\frac{V^2}{c^2}\simeq \frac{V_{\rm ej}^2}{c^2}\left(\frac{R}{r_{\rm a}}\right)^{{n-3}} \propto t^{\frac{2(n-3)}{5-n}},
\end{equation}
where the last line follows from the expression for $R(t)$ from the Sedov-Taylor solution (Equation \ref{RST}). Thus, for $n = 0$, the relativistic corrections fall off as $\propto t^{-6/5}$, and become -- as one would anticipate -- less important as the shock decelerates owing to the entrainment of mass from the ambient medium. Note, however, that this power-law decline in the importance of relativistic effects is shallower than that predicted from the Blandford-McKee solution alone, being $\Gamma \propto t^{-3/2}$ for $n = 0$ (Equation \ref{Gammaoft}); relativistic corrections to non-relativistic shock propagation can therefore be longer lived than might be anticipated by extrapolating the Blandford-McKee solution to the limit of $\Gamma =1$. Furthermore, as the density of the ambient medium falls off more steeply, relativistic effects remain important for longer periods of time (indeed, in the limit that $n \rightarrow 3$, the relativistic corrections are permanent modifications to the shock speed).

While these order of magnitude estimates provide useful diagnostics for probing the importance of relativistic effects and the rate at which they should depreciate in time, they cannot be used, for example, to determine the constant $\sigma$ that enters into the correction for the shock velocity in Equation \eqref{Vapp}. Furthermore, while these simple estimates indicate that the corrections to the post-shock fluid variables (the velocity, density, and pressure) enter at the level $V^2/c^2$, they tell us nothing about the \emph{spatial} dependence of these corrections. To understand these aspects of the problem, we now turn to a quantitative analysis of the relativistic fluid equations and treat the leading-order, relativistic corrections in the perturbative limit.

\section{Fluid Equations and Perturbative Approach}
\label{sec:fluid}
\subsection{General continuity equations}
The equations of hydrodynamics in covariant and differential form are given by

\begin{equation}
\nabla_{\mu}T^{\mu\nu} = 0, \label{enmom}
\end{equation}
where $\nabla_{\mu}$ is the covariant derivative and 

\begin{equation}
T^{\mu\nu} = \left(\rho'+\frac{\gamma}{\gamma-1}p'\right)U^{\mu}U^{\nu}+p'g^{\mu\nu}
\end{equation}
is the energy-momentum tensor of the fluid; here primes denote quantities measured in the comoving frame (i.e., in the frame where the fluid is instantaneously at rest), $\rho'$ is the fluid density, $p'$ is the gas pressure, $U^{\mu}$ is the fluid four-velocity, and $g^{\mu\nu}$ is the inverse of the metric, and we also adopted an adiabatic equation of state such that the internal energy $e'$ is related to the pressure $p'$ via $e' = p'/(\gamma-1)$. In this equation and for the remainder of this paper we adopt the Einstein summation convention, meaning that repeated upper and lower indices imply summation over the repeated index, and we will also use units where the speed of light is equal to one. The continuity equation, which ensures the conservation of mass, is written in covariant form as

\begin{equation}
\nabla_{\mu}\left[\rho'U^{\mu}\right] = 0, \label{cont}
\end{equation}
and to maintain the invariance of the line element the contraction of the four velocity with itself must be conserved:

\begin{equation}
U_{\mu}U^{\mu} = -1. \label{contract}
\end{equation}
Equations \eqref{enmom} -- \eqref{contract} constitute the inviscid conservation laws that govern the evolution of the components of the four velocity, the mass density, and the pressure of the fluid in any arbitrary geometry.

While one can deal directly with the individual components of Equation \eqref{enmom}, two additional, useful equations are obtained by contracting this expression with $U_{\nu}$ and $\Pi^{\beta}_{\,\,\nu} = U^{\beta}U_{\nu}+g^{\beta}_{\,\,\nu}$, which respectively select out the time-like and space-like components of the equations (note that $U_{\beta}\Pi^{\beta}_{\,\,\nu} = 0$). Doing so and performing some simple manipulations gives the entropy equation

\begin{equation}
U^{\mu}\nabla_{\mu}s' = 0,
\end{equation}
where $s' = \ln\left[p'/(\rho')^{\gamma}\right]$ is the entropy, and the momentum equations

\begin{equation}
\rho'U^{\mu}\nabla_{\mu}U^{\nu}+\frac{\gamma}{\gamma-1}p'U^{\mu}\nabla_{\mu}U^{\nu}+\Pi^{\mu\nu}\nabla_{\mu}p' = 0.
\end{equation}

\subsection{Equations in spherical symmetry}
Here we restrict solutions to the fluid equations to be spherically symmetric and irrotational, such that the metric is given by $g_{\mu\nu} = \text{diag}\left\{-1,1,r^2,r^2\sin^2\theta\right\}$, where $r$ is radial distance from the origin and $\theta$ is the standard polar angle in spherical coordinates, and the only non-vanishing components of the four-velocity are $U^{r} \equiv U$ and $U^{t}$. From Equation \eqref{cont} we can relate the time component of the four-velocity to the radial component via

\begin{equation}
U^{t} = \sqrt{1+U^2},
\end{equation}
and using this expression the continuity, entropy, and radial momentum equations respectively become

\begin{equation}
\frac{\partial}{\partial t}\left[\rho'\sqrt{1+U^2}\right]+\frac{1}{r^2}\frac{\partial}{\partial r}\left[r^2\rho'U\right] = 0, \label{contr}
\end{equation}
\begin{equation}
\sqrt{1+U^2}\frac{\partial s'}{\partial t}+U\frac{\partial s'}{\partial r} = 0, \label{entr}
\end{equation}
\begin{equation}
\sqrt{1+U^2}\frac{\partial U}{\partial t}+U\frac{\partial U}{\partial r}+\frac{p'U}{\rho'}\left\{\sqrt{1+U^2}\frac{\partial}{\partial t}\ln\left(p'U^{\frac{\gamma}{\gamma-1}}\right)+U\frac{\partial}{\partial r}\ln\left(p'U^{\frac{\gamma}{\gamma-1}}\right)\right\}+\frac{1}{\rho'}\frac{\partial p'}{\partial r} = 0. \label{bernr}
\end{equation}
Two other equations that will be useful in the following sections are the total energy and total radial momentum equations, which are given by the $\nu = t$ and $\nu = r$ components of Equation \eqref{enmom}. These respectively read

\begin{equation}
\frac{\partial}{\partial t}\left[\left(\rho'+\frac{\gamma}{\gamma-1}p'\right)\left(1+U^2\right)-p'\right]+\frac{1}{r^2}\frac{\partial}{\partial r}\left[r^2\left(\rho'+\frac{\gamma}{\gamma-1}p'\right)U\sqrt{1+U^2}\right]=0, \label{entot}
\end{equation}
\begin{equation}
\frac{\partial}{\partial t}\left[\left(\rho'+\frac{\gamma}{\gamma-1}p'\right)U\sqrt{1+U^2}\right]+\frac{1}{r^2}\frac{\partial}{\partial r}\left[r^2\left(\left(\rho'+\frac{\gamma}{\gamma-1}p'\right)U^2+p'\right)\right] -\frac{2p'}{r} = 0. \label{rmomtot}
\end{equation}

\subsection{Strong shock jump conditions}
\label{sec:jump}
\begin{figure}[htbp] 
   \centering
   \includegraphics[width=0.495\textwidth]{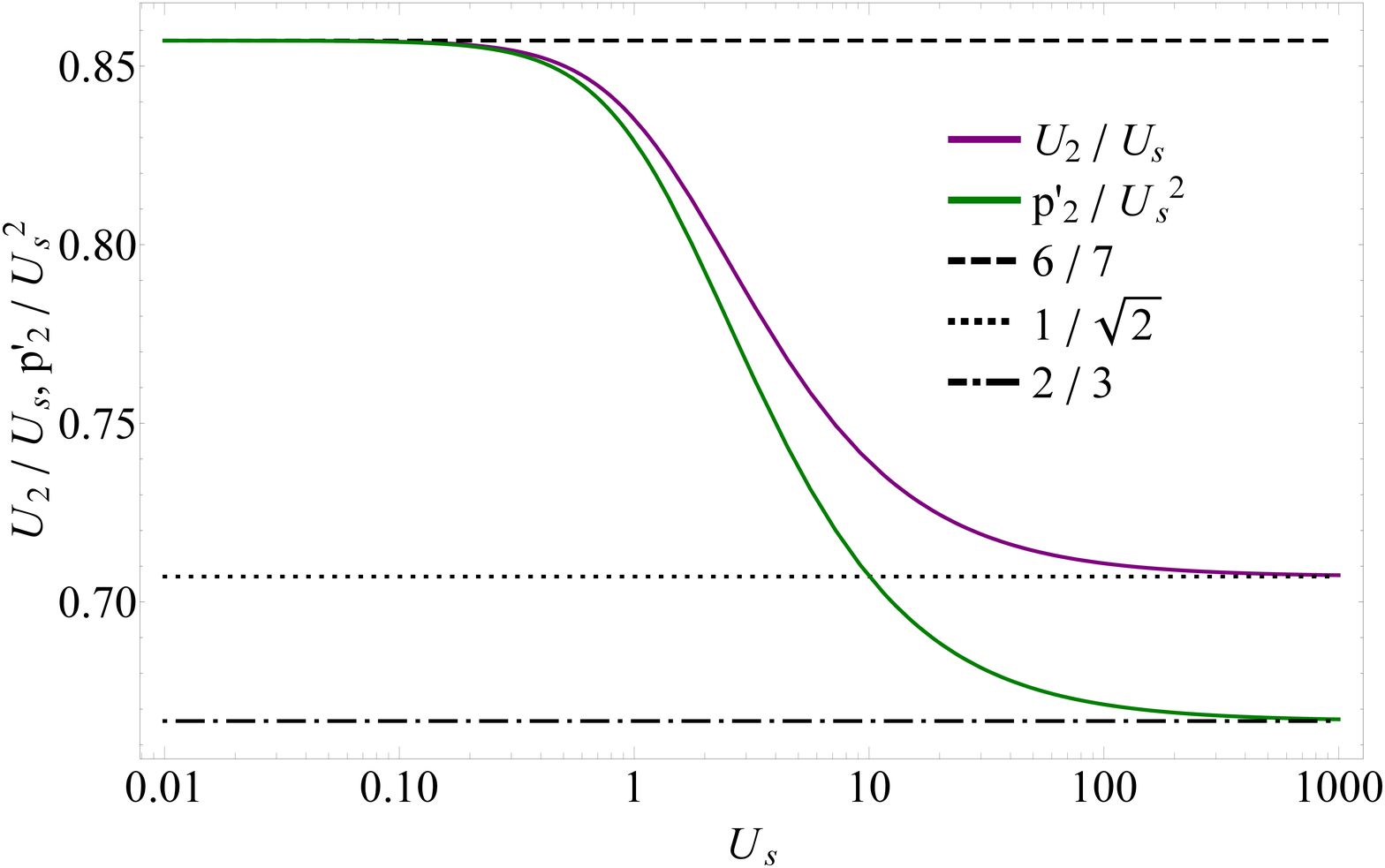} 
   \includegraphics[width=0.495\textwidth]{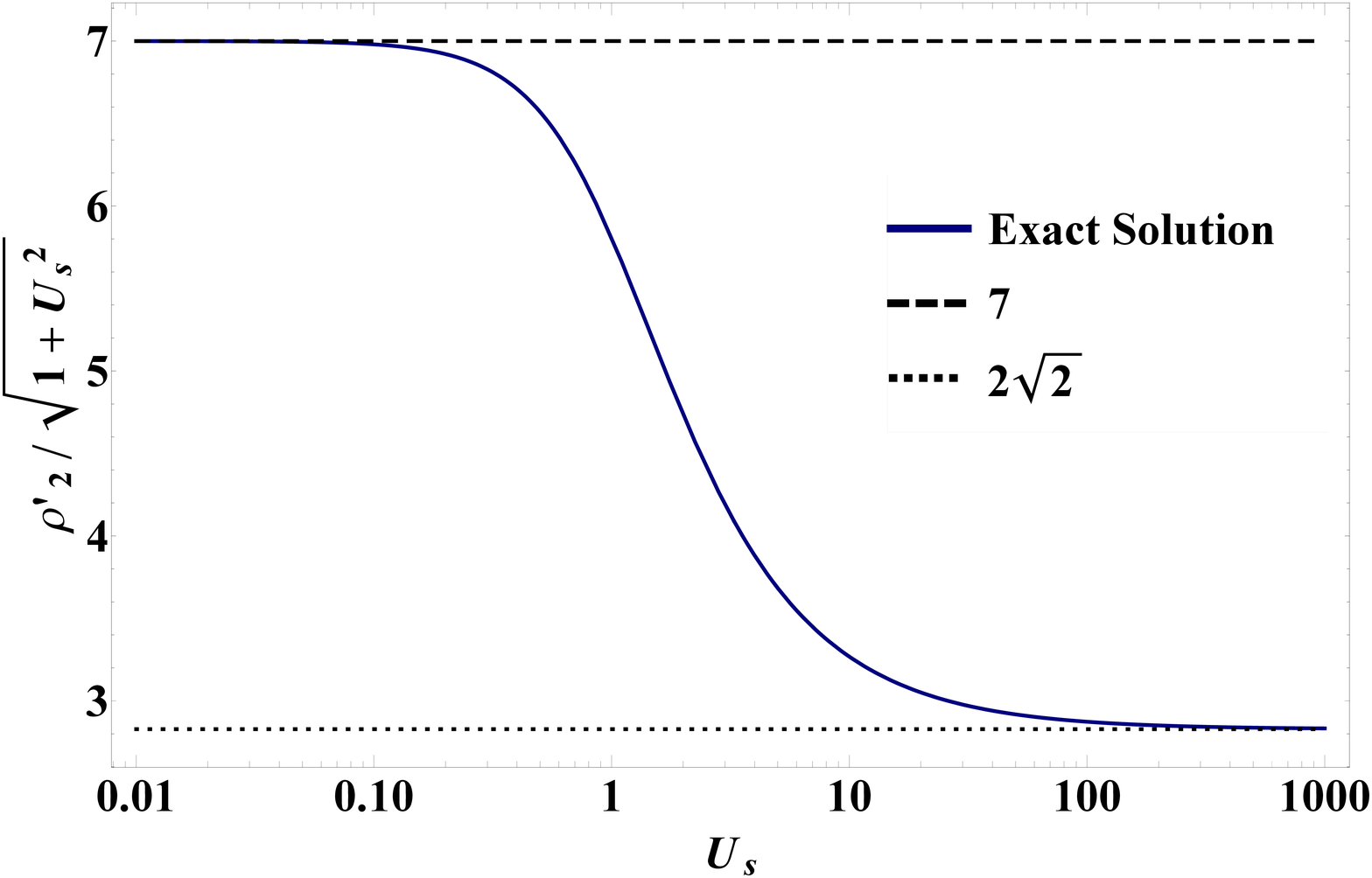} 
   \caption{Left: The exact solution for the ratio of the post-shock, lab-frame velocity to the lab-frame shock four-velocity (purple), and the exact solution for the comoving pressure normalized by the square of the lab-frame shock velocity (green) as functions of the lab-frame shock four-velocity when $\gamma = 4/3$. The non-relativistic limit of both the velocity and pressure of $6/7 \simeq 0.86$ is shown by the dashed, horizontal line, the ultra-relativistic limit of the velocity of $U_{\rm s}/\sqrt{2} \simeq 0.71$ is given by the dotted, horizontal line, and the ultra-relativistic limit of the pressure of $2U_{\rm s}^2/3$ is shown by the dot-dashed, horizontal line. Right: The exact solution for the comoving density of the fluid normalized by the Lorentz factor of the shock (blue) as a function of the lab-frame shock four-velocity when $\gamma = 4/3$. The non-relativistic limit of 7 is shown by the horizontal, dashed line, and the ultra-relativistic limit of $2\sqrt{2}$ is given by the horizontal, dotted line.}
   \label{fig:post_shock_velocity}
\end{figure}

Equations \eqref{contr} -- \eqref{bernr} govern the evolution of the post-shock fluid. Assuming that the shock generates neither mass, energy, nor momentum, the fluxes of these quantities must be conserved across the shock in the comoving frame of the shock itself. From Equations \eqref{contr}, \eqref{entot}, and \eqref{rmomtot}, conserving these fluxes yields the following three jump conditions:

\begin{equation}
\rho'_{\rm 2} U''_{\rm 2} = \rho'_{\rm a}U''_{\rm a}, \label{contsh}
\end{equation}
\begin{equation}
\left(\rho'_{\rm 2}+\frac{\gamma}{\gamma-1}p'_{\rm 2}\right)U''_{\rm 2}\sqrt{1+\left(U''_{\rm 2}\right)^2} = \left(\rho'_{\rm a}+\frac{\gamma}{\gamma-1}p'_{\rm a}\right)U''_{\rm a}\sqrt{1+\left(U''_{\rm a}\right)^2}
\end{equation}
\begin{equation}
\left(\rho'_{\rm 2}+\frac{\gamma}{\gamma-1}p'_{\rm 2}\right)\left(U''_{\rm 2}\right)^2+p'_{\rm 2} = \left(\rho'_{\rm a}+\frac{\gamma}{\gamma-1}p'_{\rm a}\right)\left(U''_{\rm a}\right)^2+p'_{\rm a}. \label{rmomsh}
\end{equation}
Here $U''$ denotes the radial component of the four-velocity in the comoving frame of the shock, quantities with a subscript 2 are post-shock fluid quantities, and those with a subscript ``a'' pertain to the ambient medium. We will assume that the shock is sufficiently supersonic that the ambient pressure can be ignored, which is equivalent to stating that the energy behind the blastwave is much greater than the internal energy of the ambient medium. In this case, these three equations can be combined into the following cubic to be solved for $U''_2$ in terms of $U''_{\rm a}$:

\begin{equation}
\gamma^2\left(1+\left(U''_{\rm 2}\right)^2\right)\left(U_{\rm 2}''-U_{\rm a}''\right) +2\gamma U''_{\rm a}\left(1+\left(U''_{\rm 2}\right)^2\right)-U''_{\rm 2}-U''_{\rm a} = 0. \label{cubic}
\end{equation}
Of the three algebraic solutions to this equation, only one is purely real for all $U_{\rm a}''$ and hence physical. If we denote the radial component of the lab-frame four-velocity of the shock by $U_{\rm s}$, then from Lorentz transformations it follows that $U''_{\rm a} = -U_{\rm s}$, while the radial component of the lab-frame post-shock velocity $U_{\rm 2}$ is given by

\begin{equation}
U_2 = U_{\rm s}\sqrt{1+\left(U''_{\rm 2}\right)^2}+U''_2\sqrt{1+U_{\rm s}^2}. \label{u2}
\end{equation}
With the solution for the comoving, post-shock fluid velocity in terms of the lab-frame velocity of the shock from Equation \eqref{cubic}, Equation \eqref{u2} yields the lab-frame post-shock fluid velocity for arbitrary shock speeds. In the nonrelativistic limit where $U_{\rm s} \ll 1$, the full expression reduces to $U_{2}/U_{\rm s} = 2/(\gamma+1)$, while in the ultra-relativistic limit where $1+U_{\rm s}^2 \simeq U_{\rm s}^2$ it becomes $U_{2}/U_{\rm s} = \sqrt{2/\gamma-1}$. Figure \ref{fig:post_shock_velocity} shows the exact solution for the ratio $U_{2}/U_{\rm s}$ (purple, solid) when $\gamma = 4/3$, the non-relativistic limit of 6/7 (black, dashed), and the ultra-relativistic limit of $1/\sqrt{2}$ (black, dotted). 

The post-shock comoving density is simply found by inverting Equation \eqref{contsh}

\begin{equation}
\rho'_{2} = -\frac{U_{\rm s}}{U''_{2}}\rho'_{\rm a}, \label{rho2}
\end{equation}
while the pressure is obtained by rearranging Equation \eqref{rmomsh} and using Equation \eqref{rho2} to remove the dependence on $\rho'_{2}$:

\begin{equation}
p'_{2} = -\frac{\gamma-1}{\gamma}\frac{U_{\rm s}}{U''_{2}}\left(\frac{\sqrt{1+U^2_{\rm s}}}{\sqrt{1+\left(U''_2\right)^2}}-1\right)\rho'_{\rm a}. \label{p2}
\end{equation}
As for the post-shock velocity, these jump conditions reduce to the known non-relativistic and ultra-relativistic limits when $U_{\rm s} \ll 1$ and $U_{\rm s} \gg 1$. The solid, green curve in the left panel of Figure \ref{fig:post_shock_velocity} shows the full solution for the post-shock pressure normalized by $U_{\rm s}^2$, and the horizontal, dot-dashed line gives the ultra-relativistic limit of $2-\gamma = 2/3$ when $\gamma = 4/3$ (the dashed, horizontal line shows the non-relativistic limit of $p'/U_{\rm s}^2= 2/(\gamma+1)= 6/7$). The right panel of Figure \ref{fig:post_shock_velocity} illustrates the full solution for the comoving density normalized by the Lorentz factor of the shock (solid, blue curve) when $\gamma = 4/3$, the dashed, horizontal line shows the non-relativistic limit of $(\gamma+1)/(\gamma-1) = 7$, and the dotted line depicts the ultra-relativistic limit of $\sqrt{2\gamma-\gamma^2}/(\gamma-1) = 2\sqrt{2}$. 

In this paper we are interested in the consequences of mildly relativistic velocities on the propagation of a strong shock and the post-shock fluid, i.e., where $U_{\rm s}/c \lesssim few$ (recall that $U_{\rm s}$ is the radial component of the four-velocity, not the three-velocity). In this case, we can Taylor expand our expressions for the post-shock fluid quantities to the lowest, non-zero order in $U_{\rm s}$ beyond the non-relativistic limit. Doing so, we find that Equations \eqref{u2} -- \eqref{p2} become

\begin{equation}
U_2(R) =\frac{2}{\gamma+1}\left(1-\frac{\gamma-1}{2}\frac{\gamma^2+1}{\left(\gamma+1\right)^3}U_{\rm s}^2\right)U_{\rm s}, \label{jumpU}
\end{equation}
\begin{equation}
\rho'_{2}(R) = \frac{\gamma+1}{\gamma-1}\left(1+\frac{2\gamma}{\left(\gamma+1\right)^3}U_{\rm s}^2\right)\rho'_{\rm a}(R),
\end{equation}
\begin{equation}
p'_{2}(R) = \frac{2}{\gamma+1}\left(1-\frac{\gamma^2\left(\gamma-1\right)}{\left(\gamma+1\right)^3}U_{\rm s}^2\right)\rho'_{\rm a}(R)U_{\rm s}^2, \label{jumpp}
\end{equation}
where for clarity we included in these relations the fact that they hold at the location of the shock position, $R$. In agreement with our arguments in Section \ref{sec:general}, these expressions contain modifications to the non-relativistic jump conditions at the order $U_{\rm s}^2/c^2$.

\subsection{Relativistic, Conserved energy}
To initiate the outward motion of the fluid and the formation of the strong shock, there is an assumed-impulsive injection of a large amount of energy $E_{\rm ej}$ (large relative to the ambient internal energy and any gravitational potential energy) into the ambient medium, which for typical supernovae is on the order of $E \simeq 10^{50-52}$ erg (though for failed and very weak supernovae, the shock energy is $E_{\rm ej} \simeq 10^{47-48}$ erg or less). Owing to the fact that the ambient medium is assumed to be pressureless, there is no source of thermal energy as the shock moves out. Therefore, under the assumption that there are no sources or sinks of energy interior to the flow (or that, if there is a compact object such as a black hole or neutron star accreting matter, the binding energy drained by the compact object is small compared to the initial energy), this initial energy must be conserved as the shock propagates outward. 

It is tempting to relate this conserved energy to the integral of the energy density over the volume enclosed by the shock, i.e., 

\begin{equation}
E_{\rm tot} = 4\pi \int_{R_{\rm c}}^{R}T^{00}r^2dr = 4\pi\int_{R_{\rm c}}^{R}\left(\left(\rho'+\frac{\gamma}{\gamma-1}p'\right)\left(1+U^2\right)-p'\right)r^2dr.
\end{equation}
(The lower bound on this integral, $R_{\rm c}$, can be non-zero for certain combinations of the ambient density profile and the adiabatic index of the post-shock flow, where the solution terminates in a contact discontinuity; see Section \ref{sec:solutions} below.) However, this energy is \emph{not conserved}, because the shock sweeps up rest mass energy from the ambient medium as it moves outward. Indeed, this additional source of energy can be seen directly by integrating Equation \eqref{entot} over the volume enclosed by the shock and rearranging:

\begin{equation}
\frac{\partial E_{\rm tot}}{\partial t} = \rho'_{\rm a}R^2 V. \label{Etotint}
\end{equation}
The right-hand side is the change in the total energy budget due to the addition of rest mass energy from the ambient medium. Thus, the total energy $E_{\rm tot}$ is not equivalent to the initial, injected energy associated with the explosion. However, we see from integrating the continuity equation from $R_{\rm c}$ to $R(t)$ that

\begin{equation}
\frac{\partial M}{\partial t} = \rho'_{\rm a}R^2 V,
\end{equation}
where

\begin{equation}
M = 4\pi \int_{R_{\rm c}}^{R}\rho'\sqrt{1+U^2}r^2dr.
\end{equation}
Using this expression for the right-hand side of Equation \eqref{Etotint} and rearranging then gives

\begin{equation}
\frac{\partial E_{\rm ej}}{\partial t} = 0,
\end{equation}
where

\begin{equation}
E_{\rm ej} = 4\pi \int_{R_{\rm c}}^{R}\left(\left(\rho'+\frac{\gamma}{\gamma-1}p'\right)\left(1+U^2\right)-p'-\rho'\sqrt{1+U^2}\right)r^2dr.
\end{equation}
This energy is manifestly conserved as the shock advances into the ambient medium, and is the exact, relativistic analog of the energy behind the blast that is relatable to the initial energy of the explosion.

Including lowest-order, relativistic corrections to this integral for the conserved energy gives

\begin{equation}
E_{\rm ej} = 4\pi \int_{R_{\rm c}}^{R}\left(\frac{1}{2}\rho'U^2+\frac{1}{\gamma-1}p'\right)r^2dr+4\pi\int_{R_{\rm c}}^{R}\left(\frac{\gamma}{\gamma-1}p'U^2+\frac{1}{8}\rho'U^4\right)r^2dr. \label{eneq}
\end{equation}
The first term is the familiar, Newtonian expression for the total (kinetic plus internal) energy behind the blastwave, and the Sedov-Taylor solution exactly conserves this quantity. The second term is the lowest-order relativistic correction, and scales as $\propto U^{4}$ (recall that the pressure behind the shock, by virtue of the shock jump conditions, is comparable to the ram pressure of the shock; specifically see Equation \ref{jumpp}). These relativistic corrections to the energy imply that, if the energy is to be absolutely conserved, there must also be corrections to the shock velocity that account for these additional terms. 

\subsection{Self-similar equations and perturbed solutions to lowest relativistic order}
When $U \ll 1$, Equations \eqref{contr} -- \eqref{bernr} reduce to the well-known Euler equations in spherical symmetry. Here, however, we are interested in the corrections to these equations that are induced by mildly relativistic flow, and hence we need to account explicitly for these corrections. The Sedov-Taylor blastwave also adopts the change of variables

\begin{equation}
r\rightarrow \xi = \frac{r}{R},
\end{equation}
which removes the time dependence of the boundary conditions at the shock front and yields self-similar solutions of the form

\begin{equation}
U = U_{\rm s}f(\xi), \quad \rho' = \rho'_{\rm a}\left(\frac{R}{r_{\rm a}}\right)^{-n}g(\xi), \quad p = \rho_{\rm a}'\left(\frac{R}{r_{\rm a}}\right)^{-n}U_{\rm s}^2 h(\xi).
\end{equation}
It is also useful to introduce the time-like variable

\begin{equation}
\chi = \ln \left(\frac{R}{r_{\rm a}}\right), 
\end{equation} 
so the temporal and spatial derivatives then transform as

\begin{equation}
\frac{\partial}{\partial t} = \frac{1}{R}\frac{dR}{dt}\left(\frac{\partial}{\partial \chi}-\xi\frac{\partial}{\partial \xi}\right) = \frac{1}{R}\frac{U_{\rm s}}{\sqrt{1+U_{\rm s}^2}}\left(\frac{\partial}{\partial \chi}-\xi\frac{\partial}{\partial \xi}\right), \quad \frac{\partial}{\partial r} = \frac{1}{R}\frac{\partial}{\partial \xi},
\end{equation}
where we used the fact that the shock three-velocity is related to the four-velocity via $dR/dt = U_{\rm s}/\sqrt{1+U_{\rm s}^2}$. The continuity, entropy, and radial momentum equations then become

\begin{equation}
\frac{U_{\rm s}}{\sqrt{1+U_{\rm s}^2}}\left(\frac{\partial}{\partial \chi}\left[\rho'\sqrt{1+U^2}\right]-\xi\frac{\partial}{\partial \xi}\left[\rho'\sqrt{1+U^2}\right]\right)+\frac{1}{\xi^2}\frac{\partial}{\partial \xi}\left[\xi^2\rho'U\right] = 0, \label{contss}
\end{equation}
\begin{equation}
U_{\rm s}\frac{\sqrt{1+U^2}}{\sqrt{1+U_{\rm s}^2}}\left(\frac{\partial s'}{\partial \chi}-\xi\frac{\partial s'}{\partial \xi}\right)+U\frac{\partial s'}{\partial \xi} = 0, \label{entss}
\end{equation}
\begin{multline}
U_{\rm s}\frac{\sqrt{1+U^2}}{\sqrt{1+U_{\rm s}^2}}\left(\frac{\partial U}{\partial \chi}-\xi\frac{\partial U}{\partial \xi}\right)+U\frac{\partial U}{\partial \xi}+\frac{p' U}{\rho'}\left(U_{\rm s}\frac{\sqrt{1+U^2}}{\sqrt{1+U_{\rm s}^2}}\left(\frac{\partial}{\partial \chi}\ln\left(p'U^{\frac{\gamma}{\gamma-1}}\right)-\xi\frac{\partial}{\partial \xi}\ln\left(p'U^{\frac{\gamma}{\gamma-1}}\right)\right)+U\frac{\partial}{\partial \xi}\ln\left(p'U^{\frac{\gamma}{\gamma-1}}\right)\right) \\ 
+\frac{1}{\rho'}\frac{\partial p'}{\partial \xi} = 0. \label{rmomss}
\end{multline}
We will now expand these equations to lowest, post-Newtonian order. Before doing so, however, we note that there are exact, self-similar solutions (being the Sedov-Taylor solutions) in the Newtonian case, and hence we expect that the velocity, density, and pressure of the fluid should be approximately given by these solutions but with perturbations that are introduced from the velocity scale set by the speed of light. Moreover, investigating the shock jump conditions \eqref{jumpU} -- \eqref{jumpp}, we see that relativistic terms modify the post-shock fluid quantities at the order $U_{\rm s}^2$. We therefore seek solutions for the velocity, comoving density, and pressure that are of the form

\begin{equation}
U = U_{\rm s}\left\{f_0(\xi)+U_{\rm s}^2f_1(\xi)\right\}, \label{Uss}
\end{equation}
\begin{equation}
\rho' = \rho'_{\rm a}\left(\frac{R}{r_{\rm a}}\right)^{-n}\left\{g_0(\xi)+U_{\rm s}^2g_1(\xi)\right\},
\end{equation}
\begin{equation}
p' = \rho'_{\rm a}\left(\frac{R}{r_{\rm a}}\right)^{-n}U_{\rm s}^2\left(h_0(\xi)+U_{\rm s}^2h_1(\xi)\right\}. \label{pss}
\end{equation}
From the shock jump conditions, the functions satisfy the following boundary conditions at the shock:

\begin{equation}
f_0(1) = h_0(1) = \frac{2}{\gamma+1}, \quad g_0(1) = \frac{\gamma+1}{\gamma-1},
\end{equation}
\begin{equation}
f_1(1) = -\frac{\gamma-1}{\gamma+1}\frac{\gamma^2+1}{\left(\gamma+1\right)^3}, \quad g_1(1) = \frac{2\gamma}{\left(\gamma-1\right)\left(\gamma+1\right)^2}, \quad h_1(1) = -\frac{2\gamma^2\left(\gamma-1\right)}{\left(\gamma+1\right)^4}. \label{bcs}
\end{equation}
In addition to satisfying these boundary conditions, energy must also be globally conserved. Returning to Equation \eqref{eneq} and using these forms for the velocity, density, and pressure, the conserved energy is given by

\begin{multline}
E = {4\pi\rho'_{\rm a}r_{\rm a}^3}\left(\frac{R}{r_{\rm a}}\right)^{3-n}U_{\rm s}^2\int_{\xi_{\rm c}}^{1}\left(\frac{1}{2}g_0f_0^2+\frac{1}{\gamma-1}h_0\right)\xi^2d\xi+{4\pi\rho'_{\rm a}r_{\rm a}^3}\left(\frac{R}{r_{\rm a}}\right)^{n-3}U_{\rm s}^2 \\ 
\times U_{\rm s}^2\int_{\xi_{\rm c}}^{1}\left(\frac{1}{2}g_1f_0^2+g_0f_0f_1+\frac{1}{\gamma-1}h_1+\frac{\gamma}{\gamma-1}h_0f_0^2+\frac{1}{8}g_0f_0^4\right)\xi^2d\xi.
\end{multline}
If the second term in this expression were absent, then the energy would be conserved if

\begin{equation}
{4\pi\rho'_{\rm a}r_{\rm a}^3}\left(\frac{R}{r_{\rm a}}\right)^{3-n}U_{\rm s}^2 = E_*,
\end{equation}
where $E_*$ is a constant that scales with the total energy, and this is just the familiar velocity-radius relation for the Sedov-Taylor blastwave. However, because the second, relativistic term modifies the energy, this scaling \emph{cannot hold exactly}, as otherwise the relativistic corrections would violate energy conservation. There must therefore be relativistic corrections to the shock velocity, and energy conservation dictates that these corrections must be of the form

\begin{equation}
{4\pi\rho'_{\rm a}r_{\rm a}^3}\left(\frac{R}{r_{\rm a}}\right)^{3-n}U_{\rm s}^2 = E_*\left(1+\sigma U_{\rm s}^2\right), \label{sigma}
\end{equation}
where $\sigma$ is a dimensionless number. Inserting this expression into the above equation for the total energy and requiring that the relativistic terms cancel exactly yields

\begin{equation}
\sigma\int_{\xi_{\rm c}}^{1}\left(\frac{1}{2}g_0f_0^2+\frac{1}{\gamma-1}h_0\right)\xi^2d\xi+\int_{\xi_{\rm c}}^{1}\left(\frac{1}{2}g_1f_0^2+g_0f_0f_1+\frac{1}{\gamma-1}h_1+\frac{\gamma}{\gamma-1}h_0f_0^2+\frac{1}{8}g_0f_0^4\right)\xi^2d\xi = 0. \label{eigenint}
\end{equation}
For a given $\sigma$, the functions $f_1$, $g_1$, and $h_1$ are completely specified from the fluid equations. This fourth boundary condition, which enforces global energy conservation, will therefore only be satisfied for a certain value of $\sigma$ (for given $n$ and $\gamma$), making $\sigma$ an ``eigenvalue'' from the standpoint that it is uniquely determined by this additional, integral constraint.

The governing equations for the self-similar functions can now be obtained by inserting Equations \eqref{Uss} -- \eqref{pss} and \eqref{sigma} into Equations \eqref{contss} -- \eqref{rmomss} and keeping the lowest-order, surviving terms in $U_{\rm s}$. Doing so and performing some algebra yields the following three equations for the non-relativistic quantities:

\begin{equation}
-ng_0-\xi\frac{dg_0}{d\xi}+\frac{1}{\xi^2}\frac{d}{d \xi}\left(\xi^2g_0f_0\right) = 0, \label{g0eq}
\end{equation}
\begin{equation}
\frac{1}{2}\left(n-3\right)f_0+\left(f_0-\xi\right)f_0'+\frac{1}{g_0}\frac{dh_0}{d\xi} = 0, \label{f0eq}
\end{equation}
\begin{equation}
-3+n\gamma+(f_0-\xi)\frac{d}{d\xi}\ln\left(\frac{h_0}{g_0^{\gamma}}\right) = 0. \label{h0eq}
\end{equation}
These three equations can be integrated three times exactly, with the pressure being analytically related to the velocity and density as

\begin{equation}
h_0 = \frac{\gamma-1}{2}\frac{\xi-f_0}{\gamma f_0-\xi}g_0f_0^2. \label{h0ex}
\end{equation}

The equations for the perturbed functions are given by

\begin{equation}
-3g_1-\xi \frac{\partial g_1}{\partial \xi}+\frac{1}{\xi^2}\frac{\partial}{\partial \xi}\left[\xi^2\left(g_0f_1+f_0g_1\right)\right]=\frac{3}{2}g_0f_0^2+\frac{1}{2}\xi\frac{\partial}{\partial \xi}\left[g_0f_0^2\right]-\frac{1}{2}\frac{1}{\xi^2}\frac{\partial}{\partial \xi}\left[\xi^2f_0g_0\right], \label{cont1}
\end{equation}
\begin{equation}
\left(n-3\right)\left(\frac{h_1}{h_0}-\gamma\frac{g_1}{g_0}\right)+\left(f_0-\xi\right)\frac{\partial}{\partial \xi}\left[\frac{h_1}{h_0}-\gamma\frac{g_1}{g_0}\right]+f_1 \frac{\partial s_0}{\partial \xi} = -\sigma\left(n-3\right)-\frac{1}{2}f_0\left(1-f_0^2\right)\frac{\partial s_0}{\partial \xi}, \label{ent1}
\end{equation}
\begin{multline}
\frac{3}{2}\left(n-3\right)f_1-\xi\frac{\partial f_1}{\partial \xi}+\frac{\partial}{\partial \xi}\left[f_0f_1\right]+\frac{1}{g_0}\left(\frac{\partial h_1}{\partial \xi}-\frac{g_1}{g_0}\frac{\partial h_0}{\partial \xi}\right) \\ 
 = -\frac{\sigma}{2}\left(n-3\right)f_0-\frac{1}{2}\left(f_0^2-1\right)\left(\frac{1}{2}\left(n-3\right)f_0-\xi\frac{\partial f_0}{\partial \xi}\right)-\frac{\gamma}{\gamma-1}\frac{f_0h_0}{g_0}\left(\frac{1}{2}n-\frac{9}{2}+\frac{3}{\gamma}+\left(f_0-\xi\right)\frac{\partial}{\partial \xi}\ln\left[f_0h_0^{\frac{\gamma-1}{\gamma}}\right]\right). \label{rmom1}
\end{multline}
Equation \eqref{cont1} can be integrated, and using the boundary conditions at the shock front gives the following expression for the relativistic correction to the density:

\begin{equation}
g_1 = \frac{\frac{1}{2}f_0\left(f_0\xi-1\right)-f_1}{f_0-\xi}g_0. \label{g1ex}
\end{equation}

The integral constraint that determines the eigenvalue, Equation \eqref{eigenint}, can also be written as a fourth boundary condition on the functions $f_1$, $g_1$, and $h_1$: subtracting the continuity from the energy equation and integrating from $R_c(t)$ to $R(t)$ gives

\begin{multline}
\frac{\partial E_{\rm ej}}{\partial t} = R^2U_{\rm s}^3\rho'_{\rm a}\left(\frac{R}{r_{\rm a}}\right)^{-n}U_{\rm s}^2\xi_{\rm c}^2 \\ \times\bigg\{
\left(g_0f_0\left(f_0-\xi_{\rm c}\right)+\frac{\gamma}{\gamma-1}h_0\right)f_1+\frac{\gamma f_0-\xi_{\rm c}}{\gamma-1}h_1-\frac{1}{4}\left(\frac{1}{2}g_0f_0^2+1\right)\left(f_0-\xi_{\rm c}\right)g_0f_0^2+\left(\gamma\left(\frac{1}{2}f_0-\xi_{\rm c}\right)f_0^2+\frac{1}{2}\xi_{\rm c}\right)h_0\bigg\}, \label{bc4}
\end{multline}
where we used Equation \eqref{g1ex} to remove the dependence on $g_1$ and all of the functions are evaluated at $\xi_{\rm c}$. By virtue of Equation \eqref{h0ex}, this expression only contains a relativistic correction, and for energy-conserving solutions we therefore require that the term in braces (multiplied by $\xi_{\rm c}^2$) be equal to zero. Either this fourth boundary condition or the integral constraint \eqref{eigenint} can be used to determine $\sigma$.

\subsection{Shock position and unperturbed coordinates}
Equation \eqref{sigma} can be rearranged and integrated numerically to yield the shock position as a function of time. However, we can also use the assumed-smallness of terms of order $U_{\rm s}^2/c^2$ to decompose the shock position and velocity into their non-relativistic and relativistically-corrected parts; denoting the non-relativistic shock position and three-velocity as $R_0$ and $V_0 = dR_0/dt$ and their relativistically-perturbed counterparts as $R_1$ and $V_1$, we find 

\begin{equation}
4\pi \rho'_{\rm a} r_{\rm a}^3\left(\frac{R_0}{r_{\rm a}}\right)^{3-n}V_0^2\left(1+\left(3-n\right)\frac{R_1}{R_0}\right)\left(1+2\frac{V_1}{V_0}\right)\left(1+\frac{V_0^2}{c^2}\right) = E_*\left(1+\sigma \frac{V_0^2}{c^2}\right),
\end{equation}
where we introduced factors of $c^2$ for clarity. This expression demonstrates, as expected, that the unperturbed shock velocity and position are related via the standard, energy-conserving prescription for the Sedov-Taylor blastwave:

\begin{equation}
4\pi r_{\rm a}^3\rho_{\rm a}\left(\frac{R_0}{r_{\rm a}}\right)^{3-n}V_0^2 = E_* \quad \Rightarrow \quad \frac{R_0}{r_{\rm a}} = \left(1+\frac{5-n}{2}\frac{V_{\rm i}}{r_{\rm a}}t\right)^{\frac{2}{5-n}}, \label{unpert}
\end{equation}
where we defined $V_{\rm i}^2 = E_*/(4\pi \rho_{\rm a}r_{\rm a}^3)$ as the unperturbed velocity of the shock when the shock position coincides with $R = r_{\rm a}$, while the relativistic corrections to the shock position and velocity satisfy

\begin{equation}
\frac{V_1}{V_0}+\frac{3-n}{2}\frac{R_1}{R_0} = \frac{\sigma-1}{2}\frac{V_0^2}{c^2}.
\end{equation}
This equation can be integrated to yield, if $n \neq 1$, 

\begin{equation}
\frac{R_1}{R_0} 
= \frac{\sigma-1}{n-1}\frac{V_{\rm i}^2}{c^2}\left(\left(\frac{R_0}{r_{\rm a}}\right)^{n-3}-\left(\frac{R_0}{r_{\rm a}}\right)^{\frac{n-5}{2}}\right), \label{R1eq}
\end{equation}
\begin{equation}
\frac{V_1}{V_0} 
= \frac{\sigma-1}{n-1}\frac{V_{\rm i}^2}{c^2}\left(\left(n-2\right)\left(\frac{R_0}{r_{\rm a}}\right)^{n-3}-\frac{n-3}{2}\left(\frac{R_0}{r_{\rm a}}\right)^{\frac{n-5}{2}}\right),\label{V1eq}
\end{equation}
while if $n =1$

\begin{equation}
\frac{R_1}{R_0} = \frac{\sigma-1}{2}\frac{V_{\rm i}^2}{c^2}\left(\frac{R_0}{r_{\rm a}}\right)^{-2}\ln\left[\frac{R_0}{r_{\rm a}}\right],
\end{equation}
\begin{equation}
\frac{V_1}{V_0} = \frac{\sigma-1}{2}\frac{V_{\rm i}^2}{c^2}\left(\frac{R_0}{r_{\rm a}}\right)^{-2}\left(1-\ln\left[\frac{R_0}{r_{\rm a}}\right]\right).
\end{equation}
The second term in parentheses in Equations \eqref{R1eq} and \eqref{V1eq} is a consequence of initial conditions, and arises from the fact that scale invariance allows us to define the relativistic corrections to the shock position to be zero at $t = 0$. Interestingly, if $\sigma \equiv 1$, then the relativistic corrections to the shock velocity and position are exactly zero. This effect arises from a competition between the increase in the four-velocity generated by positive $\sigma$, and time dilation that reduces the three-velocity from the four-velocity -- when $\sigma = 1$ these two effects exactly balance to yield no relativistic correction to the shock velocity. 

Following \citet{coughlin19}, we wrote our solutions for the relativistic corrections to the velocity, density, and pressure of the post-shock fluid in terms of the true shock position and velocity and the total self-similar variable $\xi = r/R(t)$. While formally correct to order $U_{\rm s}^2/c^2$, these expressions (specifically Equations \ref{Uss} -- \ref{pss}) also contain terms that are of a higher order than $U_{\rm s}^2/c^2$. We can remove these additional terms by rewriting the solutions in terms of the ``unperturbed'' self-similar variable $\xi_0 = r/R_0(t)$ and using Equations \eqref{R1eq} and \eqref{V1eq} to write the corrections to the shock position and velocity in terms of their non-relativistic counterparts; the resulting expressions, which are identical to Equations \eqref{Uss} -- \eqref{pss} to order $U_{\rm s}^2/c^2$, are

\begin{equation}
U(\xi_0,t) = V_0 \left(f_0(\xi_0)+\left(\frac{V_1}{V_0}+\frac{1}{2}V_0^2\right)f_0(\xi_0)-\xi_0 f_0'(\xi_0)\frac{R_1}{R_0}+V_0^2f_1(\xi_0)\right) \equiv V_0\left\{f_0(\xi_0)+f_1^*(\xi_0,t)\right\},
\end{equation}
\begin{equation}
\rho'(\xi_0,t) = \rho'_{\rm a}\left(\frac{R_0}{r_{\rm a}}\right)^{-n}\left(g_0(\xi_0)-\left(ng_0+\xi_0g_0'\right)\frac{R_1}{R_0}+V_0^2g_1(\xi_0)\right) \equiv \rho'_{\rm a}\left(\frac{R_0}{r_{\rm a}}\right)^{-n}\left\{g_0(\xi_0)+g_1^*(\xi_0,t)\right\},
\end{equation}
\begin{multline}
p'(\xi_0,t) = \rho'_{\rm a}\left(\frac{R_0}{r_{\rm a}}\right)^{-n}V_0^2\left(h_0(\xi_0)-\left(nh_0+\xi h_0'\right)\frac{R_1}{R_0}+2\left(\frac{V_1}{V_0}+\frac{1}{2}V_0^2\right)h_0(\xi_0)+V_0^2h_1(\xi_0)\right) \\ 
 \equiv \rho'_{\rm a}\left(\frac{R_0}{r_{\rm a}}\right)^{-n}V_0^2\left\{h_0(\xi_0)+h_1^*(\xi_0,t)\right\}. \label{presslab}
\end{multline}
Finally, the three-velocity of the fluid is

\begin{equation}
v = \frac{U}{\sqrt{1+U^2}}\simeq V_0\left\{f_0(\xi_0)+f_1^*(\xi_0,t)-\frac{1}{2}V_0^2f_0(\xi_0)^3\right\}, \label{threevel}
\end{equation}
and the lab-frame density is given by

\begin{equation}
\rho = \rho'\sqrt{1+U^2} \simeq \rho'_{\rm a}\left(\frac{R_0}{r_{\rm a}}\right)^{-n}\left\{g_0(\xi_0)+g_1^*(\xi_0,t)+\frac{1}{2}V_0^2f_0(\xi_0)^2g_0(\xi_0)\right\}. \label{labrho}
\end{equation}

\section{Solutions}
\label{sec:solutions}
Here we present the numerical solutions for the functions $f_1$, $g_1$, $h_1$, and the eigenvalue $\sigma$ that satisfy the differential Equations \eqref{ent1} and \eqref{rmom1}, with Equation \eqref{g1ex} relating $g_1$ to $f_1$, and the fourth, energy-conserving boundary condition, given either by the integral constraint \eqref{eigenint} or Equation \eqref{bc4}. Since the functions $f_1$, $g_1$, and $h_1$ satisfy the boundary conditions at the shock front, given by Equation \eqref{bcs}, we can numerically integrate Equations \eqref{ent1} and \eqref{rmom1} from $\xi = 1$ for a given, initial guess for $\sigma$. We then perturb the guess for $\sigma$ and calculate the change in the energy residual, i.e., we determine how well the new value of $\sigma$ satisfies the fourth boundary condition given by Equation \eqref{eigenint} or \eqref{bc4}, which motivates the choice for the next $\sigma$ that will better satisfy the fourth boundary condition. In this way, we iteratively determine the eigenvalue $\sigma$ that globally conserves the energy behind the blastwave. 

\begin{figure}[htbp] 
   \centering
   \includegraphics[width=0.495\textwidth]{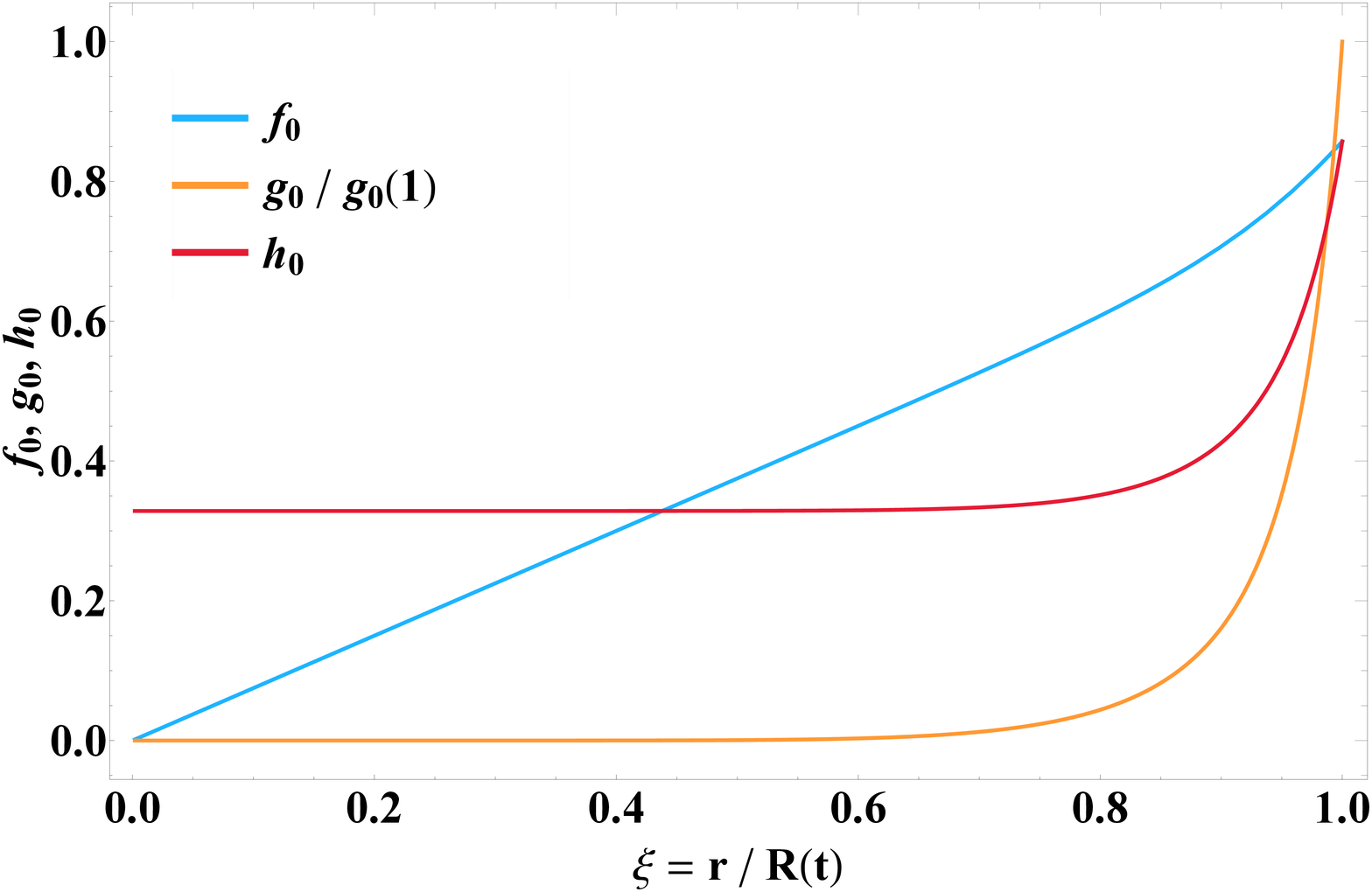} 
   \includegraphics[width=0.495\textwidth]{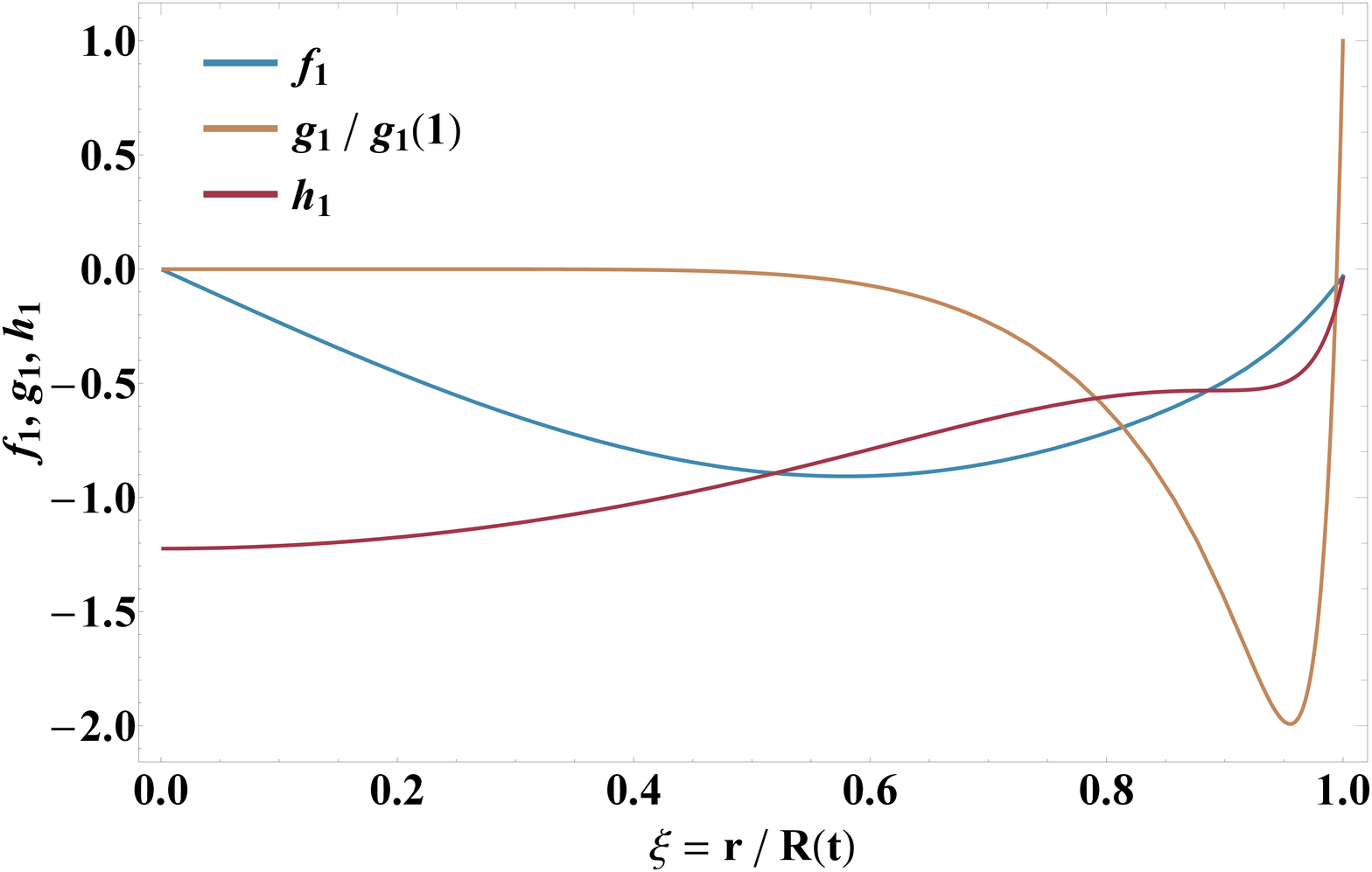} 
   \caption{Left: The Sedov-Taylor, self-similar velocity (blue), which is the four-velocity of the fluid normalized by the shock speed; the density (orange), which is the comoving density normalized by the ambient density; and the pressure (red), which is the gas pressure normalized by the ram pressure of the shock. Here we set $\gamma = 4/3$ and $n = 0$, such that the post-shock gas is radiation-pressure dominated and the ambient medium has a constant density. These fluid variables are plotted as functions of the self-similar variable $\xi$, which is just the spherical radius $r$ normalized by the shock position at a given time. The pressure is almost exactly constant, the velocity is effectively linear, and the density falls off extremely rapidly near the origin. Right: The self-similar, relativistic correction to the fluid four-velocity (dark blue), comoving density (dark orange), and gas pressure (dark red) when $\gamma = 4/3$ and $n = 0$. These solutions satisfy global energy conservation and the jump conditions at the shock, and the eigenvalue that ensures energy conservation is $\sigma \simeq 1.078$. The fact that the density is positive very near the shock but then becomes negative indicates that relativistic effects push more mass toward the shock, and the variation in the the self-similar pressure implies that the total pressure (i.e., including the relativistic terms) is less homogeneous than in the Newtonian, purely self-similar limit (left panel).}
   \label{fig:f0g0h0_n0_g43}
\end{figure}

\begin{figure}[htbp] 
   \centering
      \includegraphics[width=0.495\textwidth]{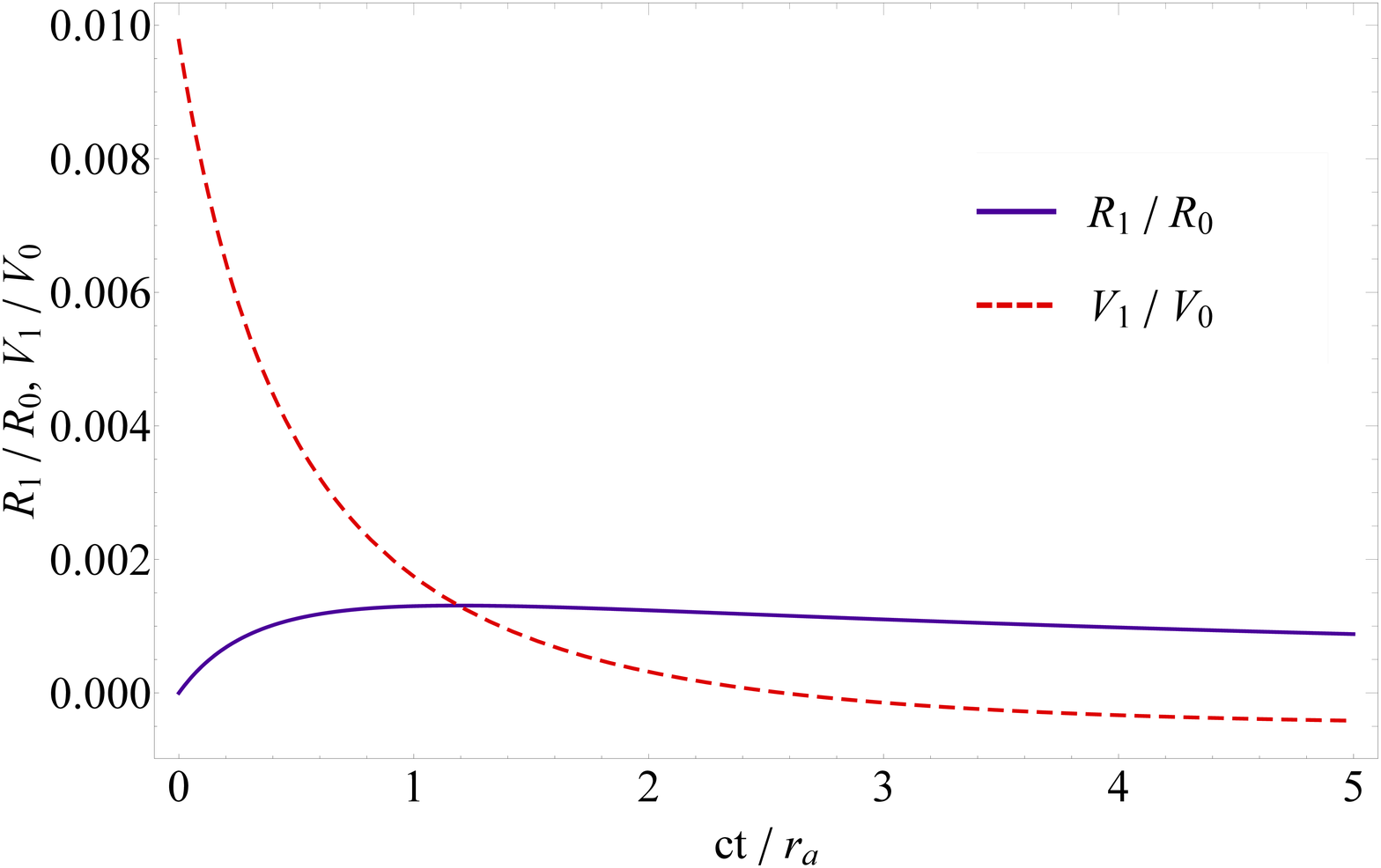} 
   \includegraphics[width=0.495\textwidth]{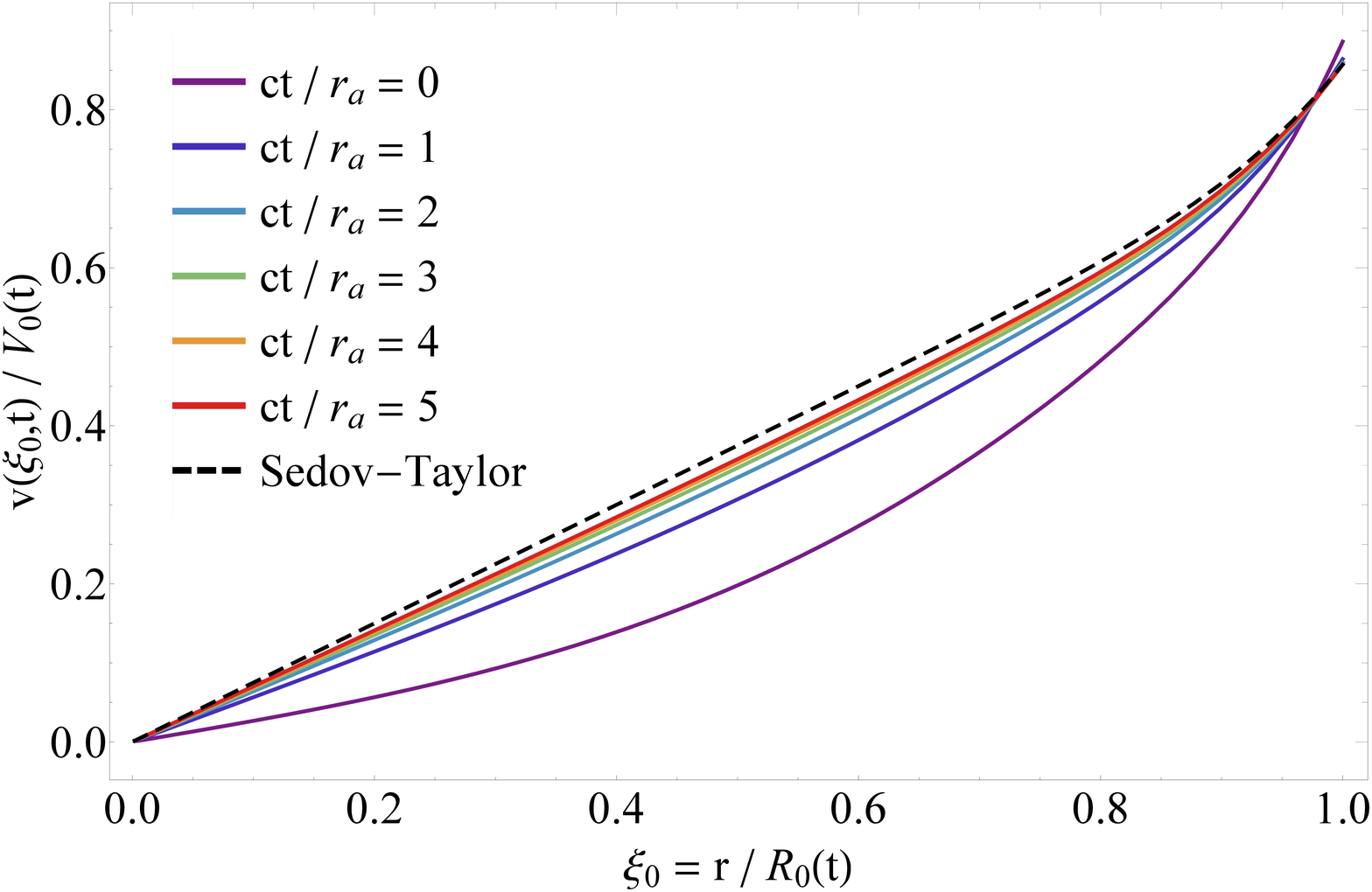} 
   \caption{Left: The relativistic correction to the shock position (purple, solid) and the shock velocity (red, dashed), both normalized by their non-relativistic counterparts, for a constant ambient density ($n = 0$) and a post-shock adiabatic index of $\gamma = 4/3$. Here we set the initial velocity of the shock to $V_{\rm i}/c = 0.5$. Right: The three-velocity of the fluid, normalized by the unperturbed shock velocity, as a function of radius $r$ normalized to the unperturbed shock radius, for $n = 0$ (constant ambient density), $\gamma = 4/3$, and an initial shock velocity of $V_{\rm i}/c = 0.5$. The different curves are at the times shown in the legend, and the black, dashed curve shows the Sedov-Taylor solution -- which would be \emph{the solution} if there were no relativistic terms -- to which the relativistic solution asymptotes at late times. The post-shock speed is slightly increased near the shock front, but falls below the Sedov-Taylor solution for smaller radii. The solution near the origin also shows significant deviation from the nearly-linear behavior expected from the Sedov-Taylor solution alone.}
   \label{fig:R1_V1_n0_g43}
\end{figure}

The left panel of Figure \ref{fig:f0g0h0_n0_g43} shows the self-similar velocity, density, and pressure for the Sedov-Taylor blastwave when $\gamma = 4/3$ and $n = 0$, corresponding to a radiation-pressure dominated, post-shock fluid and a constant ambient density. 
The linear velocity, constant pressure, and approximately zero density near the origin are familiar features of the Sedov-Taylor blastwave. The right panel of this figure gives the relativistic corrections to the velocity, density, and pressure for this combination of $n$ and $\gamma$. The eigenvalue that results in the exact conservation of energy, including the relativistic terms, is $\sigma \simeq 1.078$. The right panel of this figure also shows that, very near the shock front, the perturbation to the density is positive, and this is just a consequence of the jump conditions at the shock. However, the perturbation to the density drops very steeply from the shock front inward, and becomes negative and reaches a relative minimum near $\xi \simeq 0.95$. This qualitative behavior implies that the material behind the shock becomes increasingly confined to a region very near the shock front, and that relativistic effects cause the material to be swept into an even thinner shell than the one predicted by the Sedov-Taylor blastwave alone. We also see that the perturbation to the pressure, while it does asymptote to a constant near the origin, shows much more variability than the unperturbed solution for $\xi \gtrsim 0.1$. This feature demonstrates that the total pressure behind the blast wave shows more spatial variation when relativistic effects are included, which is a familiar property of the ultra-relativistic, Blandford-McKee blastwave.

\begin{figure}[htbp] 
   \centering
      \includegraphics[width=0.495\textwidth]{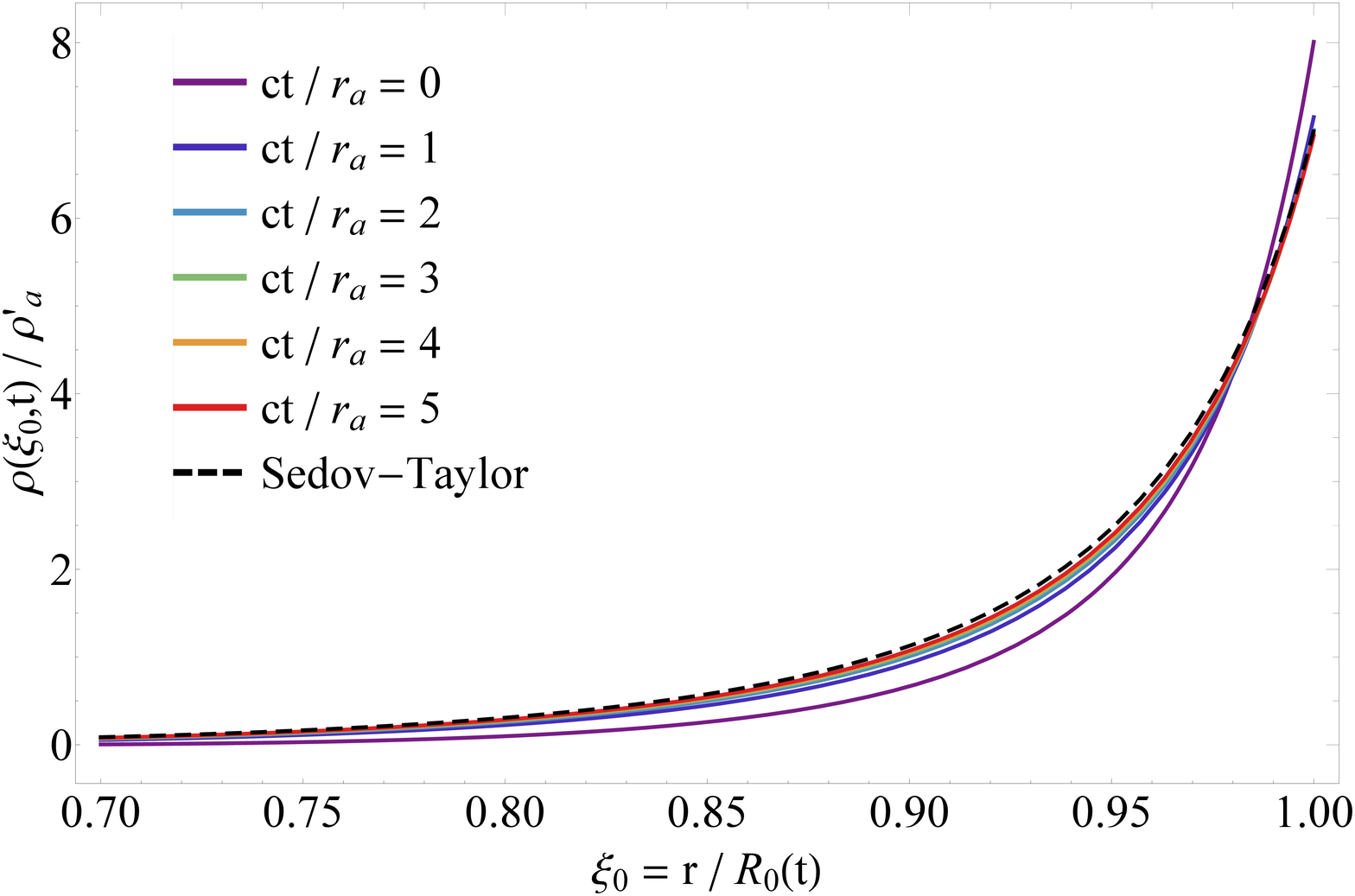} 
   \includegraphics[width=0.495\textwidth]{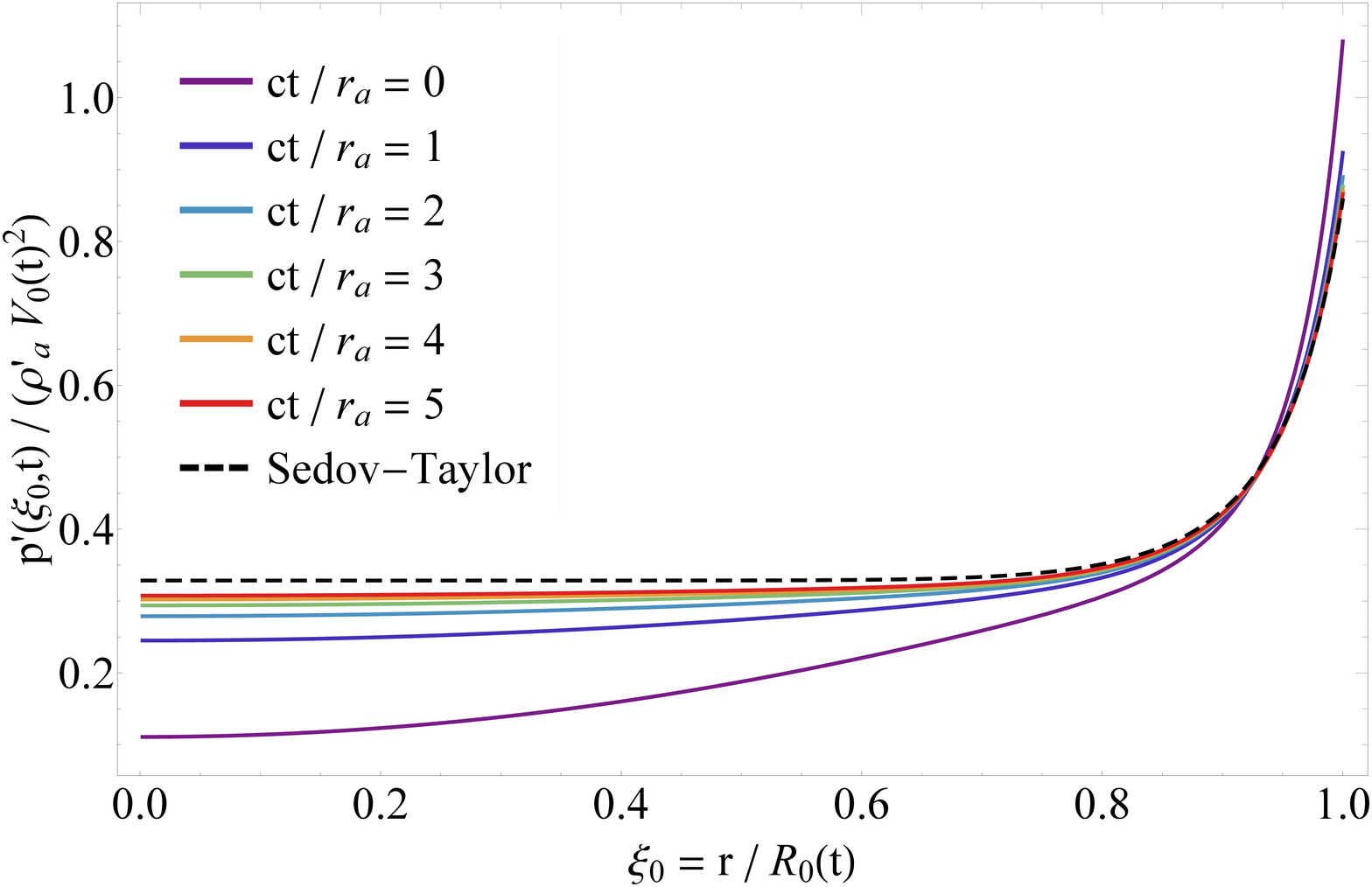} 
   \caption{Left: The lab-frame density of the post-shock fluid, normalized by the density of the ambient medium at the position of the shock $\rho'_{\rm a}(t) = \rho'_{\rm a}\left(R_0(t)/r_{\rm a}\right)^{-n}$, as a function of spherical radius $r$ normalized to the non-relativistic shock position. As for Figure \ref{fig:R1_V1_n0_g43}, here we set $n = 0$, corresponding to a constant ambient density, the post-shock adiabatic index to $\gamma = 4/3$, and the initial shock speed is $V_{\rm i}/c = 0.5$. The colored curves correspond to the times in the legend, and the black, dashed curve is the Sedov-Taylor solution. Note that, for this figure, we restricted the range of $\xi_0$ to be $\xi_0 > 0.7$, as below this range the curves all rapidly approach zero (and show little variation from one another). The density is slightly increased relative to the non-relativistic solution near the shock front, but falls below the Sedov-Taylor prediction at smaller radii, which demonstrates that the mass behind the blastwave becomes further concentrated near the shock front as the solution becomes more relativistic. Right: The pressure behind the shock for the same parameters as in the left panel. We see that relativistic effects reduce the post-shock pressure from the value predicted by the Sedov-Taylor solution near the origin, and the pressure also shows more significant variation than the nearly flat profile predicted in the non-relativistic limit. }
   \label{fig:lab_rho_n0_g43}
\end{figure}

The left panel of Figure \ref{fig:R1_V1_n0_g43} gives the correction to the shock position (solid, purple) and the correction to the shock speed (dashed, red), each normalized by its non-relativistic counterpart, as functions of time in units of $r_{\rm a}/c$. Here we set $n = 0$, corresponding to a constant ambient density, the post-shock adiabatic index to $\gamma = 4/3$, and the initial shock velocity to $V_{\rm i}/c = 0.5$. We see that the relativistic correction to the shock velocity is initially positive, corresponding to an increase in the shock position over the non-relativistic value, while at late times the correction to the shock speed changes sign; this behavior is due to the competition between the effects of positive-$\sigma$, which increases the four-velocity (see Equation \ref{sigma}), and time dilation, which reduces the three-velocity over the four-velocity. In the asymptotic limit of $t \rightarrow \infty$, both of these corrections decay to zero and the flow settles into the non-relativistic regime.

The right panel of Figure \ref{fig:R1_V1_n0_g43} shows the post-shock fluid three-velocity, normalized by the initial shock speed, as a function of normalized radial position behind the shock front (see Equation \ref{threevel}). As for the left panel of this figure, here the ambient density is constant ($n = 0$), the adiabatic index is $\gamma = 4/3$, and the initial shock speed is $V_{\rm i}/c = 0.5$. The different, colored curves correspond to the times shown in the legend, and the black, dashed curve is the Sedov-Taylor solution for this combination of $n$ and $\gamma$ (and is identical to the blue curve in the left panel of Figure \ref{fig:f0g0h0_n0_g43}). We see that near the shock front the post-shock velocity is slightly increased, which is reasonable given the slight increase in the shock velocity itself, as demonstrated in the left panel of this figure. However, at small radii the velocity falls significantly below the non-relativistic prediction, and also displays more nonlinear behavior near the origin.

The left panel of Figure \ref{fig:lab_rho_n0_g43} illustrates the normalized, lab-frame, post-shock density as a function of spherical radius $r$ normalized by the shock radius, while the right panel of this figure gives the post-shock pressure normalized by the shock velocity (see Equations \ref{labrho} and \ref{presslab} respectively). As for the right panel of Figure \ref{fig:R1_V1_n0_g43}, the different colored curves correspond to the times in the legend, the black, dashed curves are the Sedov-Taylor prediction, and we set $n = 0$ (constant ambient density), $\gamma = 4/3$, and $V_{\rm i}/c = 0.5$ (note that the $x$-axis in the left panel is compressed to highlight the behavior near the shock; for $\xi_0 \lesssim 0.7$, all of the functions rapidly approach zero and show little deviation from one another). We see that both the post-shock density and pressure increase above the Sedov-Taylor solution near the shock front, but, as is also true for the post-shock velocity, each of these quantities declines and falls below the non-relativistic prediction at a radius not far behind the shock front. The post-shock pressure also shows significant deviation from the nearly-constant value expected from the Sedov-Taylor blastwave. These findings confirm that relativistic effects tend to compress the fluid into a more confined region immediately behind the shock.

\begin{figure}[htbp] 
   \centering
   \includegraphics[width=0.495\textwidth]{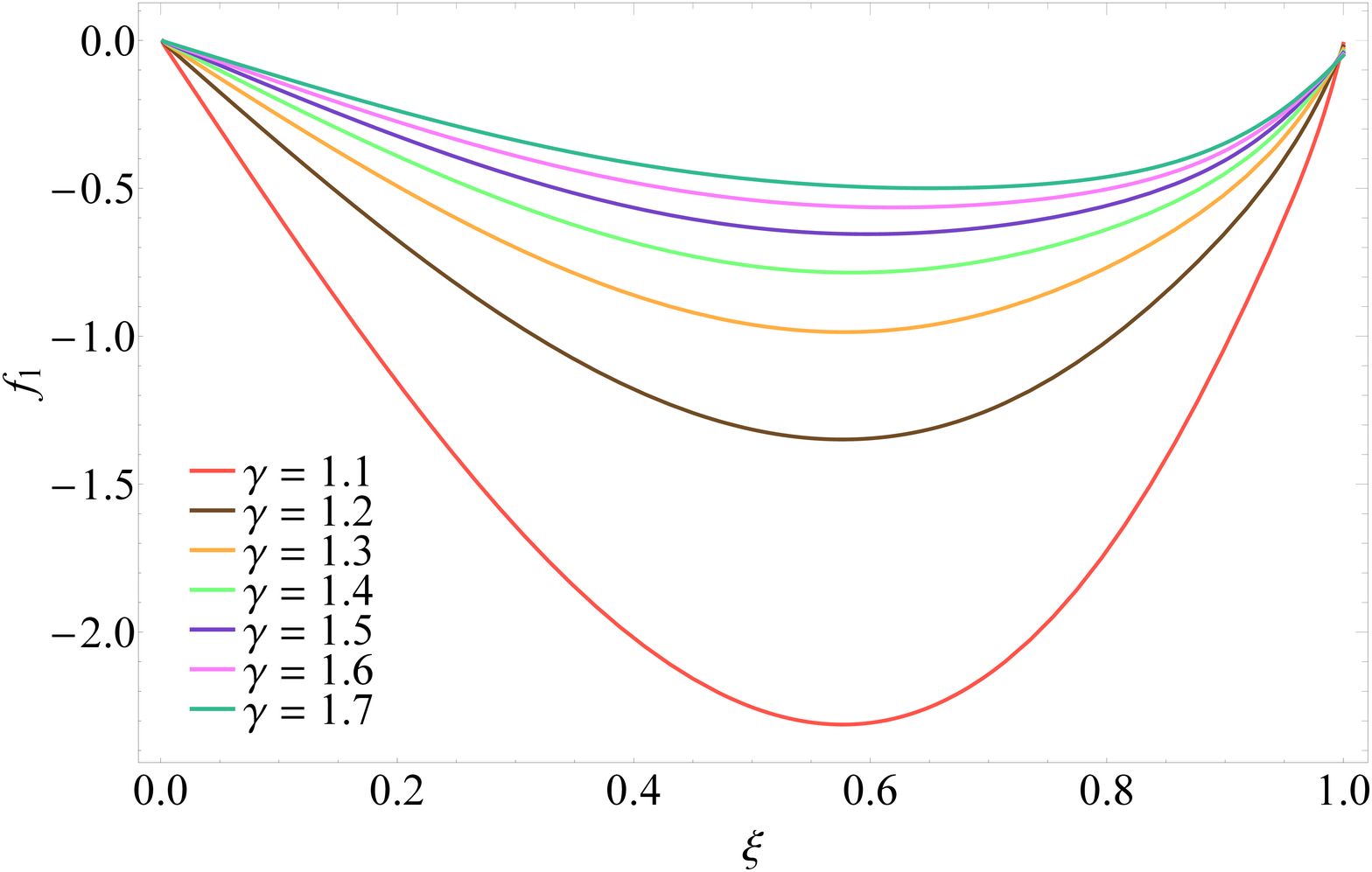} 
   \includegraphics[width=0.495\textwidth]{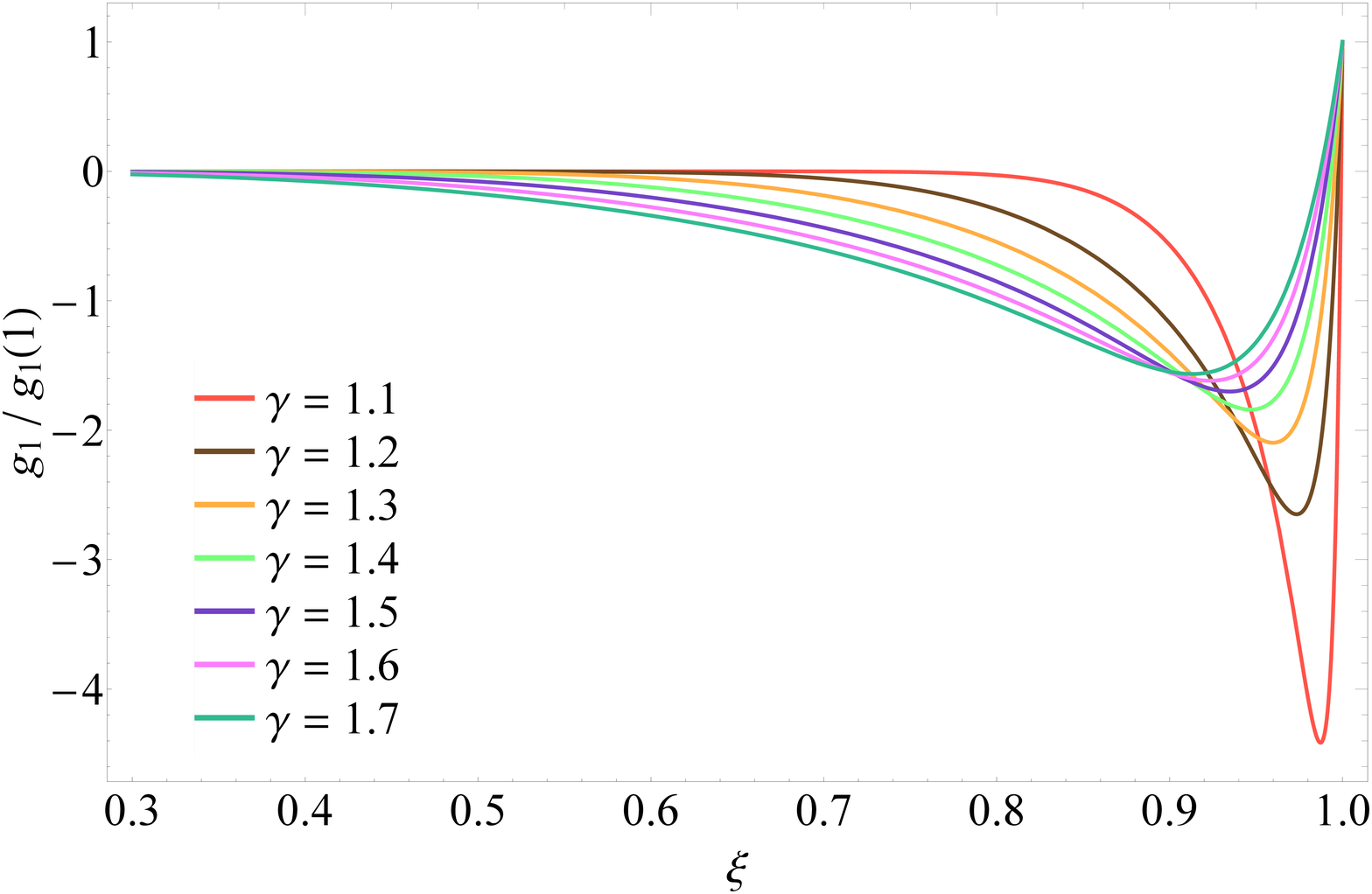} 
   \caption{Left: The self-similar, relativistic correction to the velocity when $n = 0$ -- corresponding to a constant-density ambient medium -- and the adiabatic index of the gas is given by those in the legend. Each curve shows the same, rough trend, and is negative throughout the entire post-shock flow, reaches a minimum value near $\xi \simeq 0.7$, and approaches zero near the origin; the latter feature ensures that the origin remains fixed for all of these solutions. The magnitude of the relativistic correction to the velocity grows as the adiabatic index decreases, as does the eigenvalue $\sigma$ that ensures that the solutions conserve the energy behind the blastwave (e.g., $\gamma = 1.1$ has $\sigma \simeq 3$, while $\gamma = 1.7$ has $\sigma \simeq 0.7$; see Table \ref{tab:1} for a list of eigenvalues over a range of $n$ and $\gamma$). Right: The post-shock, relativistic correction to the density for the same parameters as the left panel. As $\gamma$ approaches 1, the density becomes increasingly positive toward the shock front (note that $g_1(1) \propto (\gamma-1)^{-1}$) but also reaches an increasingly negative value, and the transition to negative values approaches the location of the shock itself. This feature demonstrates that the relativistic effects, which compress the post-shock fluid to a region that is more confined to the location of the shock itself, become more important as $\gamma$ decreases.}
   \label{fig:f1_n0_of_gamma}
\end{figure}

\begin{figure}[htbp] 
   \centering
   \includegraphics[width=0.495\textwidth]{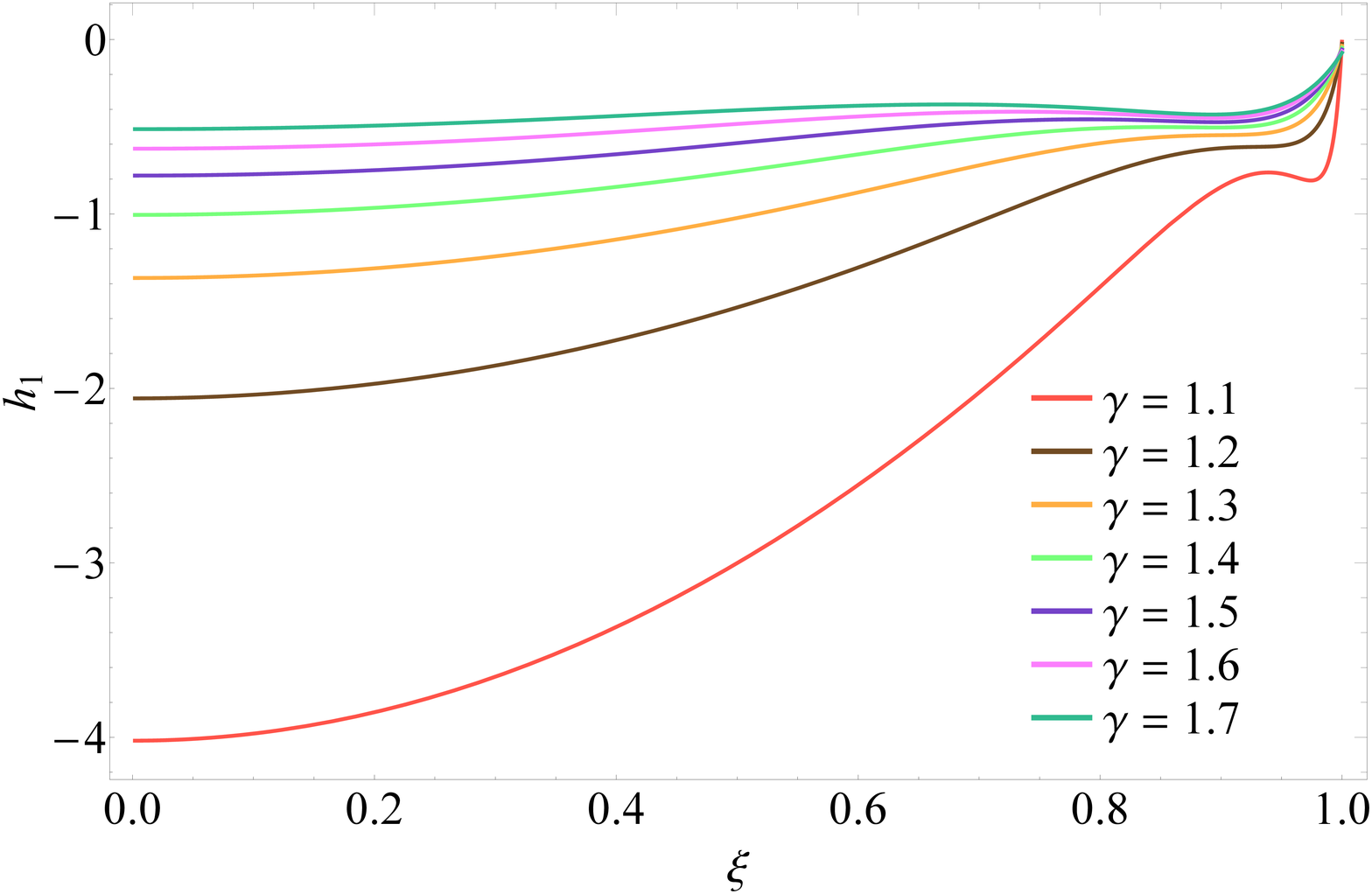} 
   \includegraphics[width=0.495\textwidth]{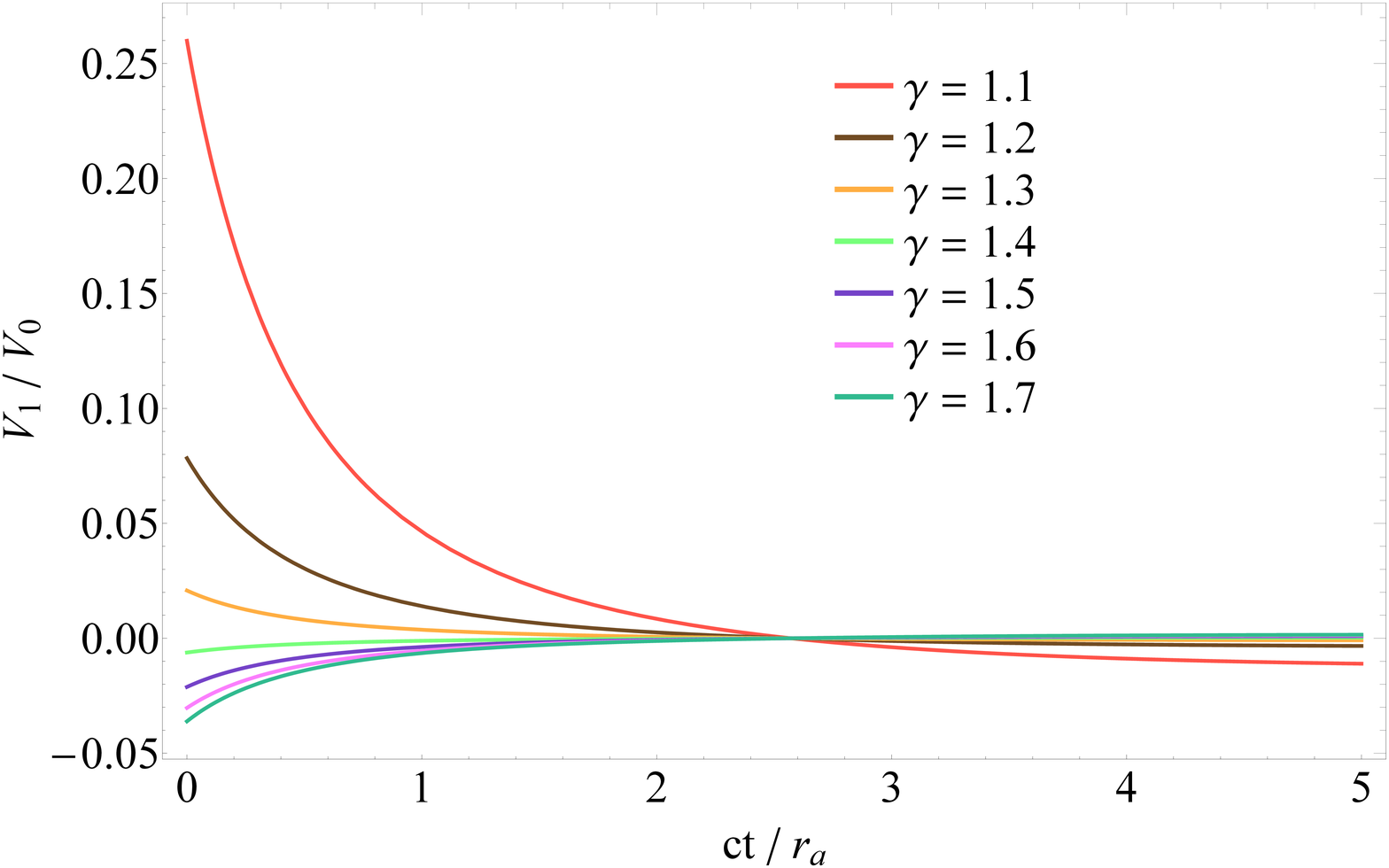} 
   \caption{Left: The post-shock, relativistic correction to the pressure behind the blastwave for a constant-density ambient medium ($n = 0$) and the range of adiabatic indices shown in the legend. As the adiabatic index decreases, the reduction in the post-shock pressure becomes more drastic near the origin, and the region over which the pressure experiences inflection points becomes more localized to the shock itself. Right: The relativistic correction to the shock velocity, normalized by the non-relativistic shock velocity, for the same parameters as the left panel; here we set the initial, unperturbed shock velocity to $V_{\rm i}/c = 0.5$. Because the eigenvalue increases as $\gamma$ decreases, the initial correction to the velocity becomes correspondingly larger. Interestingly, however, there is a value of $\gamma$ at which $\sigma$ drops below one, implying that the initial, relativistic correction to the velocity changes sign. This change in sign is due to the fact that the lab-frame speed is affected by time dilation, which can outweigh the increase in the four-velocity (which is the three-velocity in the comoving frame of the non-relativistic shock) imparted by the positive value of $\sigma$. }
   \label{fig:h1_n0_of_gamma}
\end{figure}

The left Panel of Figure \ref{fig:f1_n0_of_gamma} gives the post-shock, relativistic correction to the velocity profile of the fluid, the right panel of this figure shows the relativistic correction to the post-shock, comoving density, and the left panel of Figure \ref{fig:h1_n0_of_gamma} illustrates the post-shock correction to the pressure, and all of these panels set $n = 0$ (constant density ambient medium). The different curves in each of these figures correspond to the adiabatic indices shown in the legend. We see that, while all of these curves show the same qualitative trends, relativistic effects become amplified as the adiabatic index of the gas decreases: the magnitude of the velocity reduction is more pronounced; the material behind the shock becomes increasingly compressed to the shock itself; and the pressure has increased variation near the shock, possesses more nonlinear behavior, and the decrease near the origin is enhanced. These findings -- that relativistic effects become more important for smaller adiabatic indices -- are consistent with the fact that the eigenvalue $\sigma$ increases as $\gamma$ decreases (see Table \ref{tab:1}).

The right panel of Figure \ref{fig:h1_n0_of_gamma} shows the relativistic correction to the shock velocity, normalized by the non-relativistic shock speed, as a function of time in units of $r_{\rm a}/c$. The different curves are appropriate to the adiabatic indices shown in the legend, the ambient density profile is constant ($n = 0$), and we set the initial, non-relativistic shock speed to $V_{\rm i}/c = 0.5$. We see that, when $\gamma$ is small, the initial velocity increases compared to the non-relativistic one, but becomes negative after a time of $c t/r_{\rm a} \simeq 2.57$ (the exact time at which this occurs can be derived analytically from Equation \ref{V1eq}). However, for values of $\gamma \gtrsim 1.4$, this behavior inverts, with the initial correction to the velocity being negative at early times and transitioning to positive after $c t/r_{\rm a} \simeq 2.57$. This inversion occurs mathematically because $\sigma$ crosses $\sigma = 1$ at this location, which is where the relativistic correction to the three-velocity equals zero (this feature can be seen by inserting $\sigma = 1$ into Equation \ref{sigma}), and -- as we noted above -- arises physically because the three-velocity of the fluid is affected by time dilation. 

\begin{figure}[htbp] 
   \centering
   \includegraphics[width=0.495\textwidth]{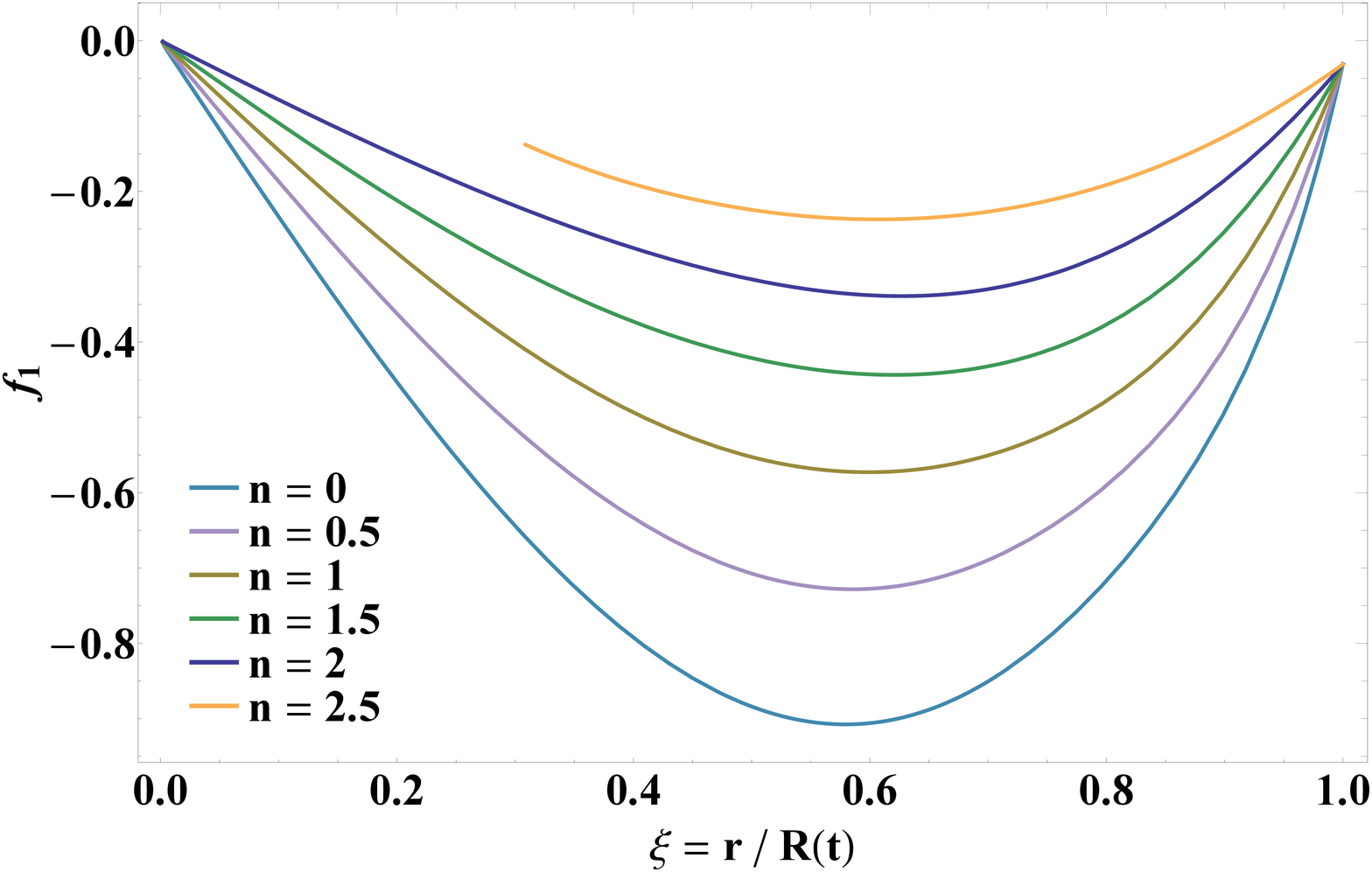} 
   \includegraphics[width=0.495\textwidth]{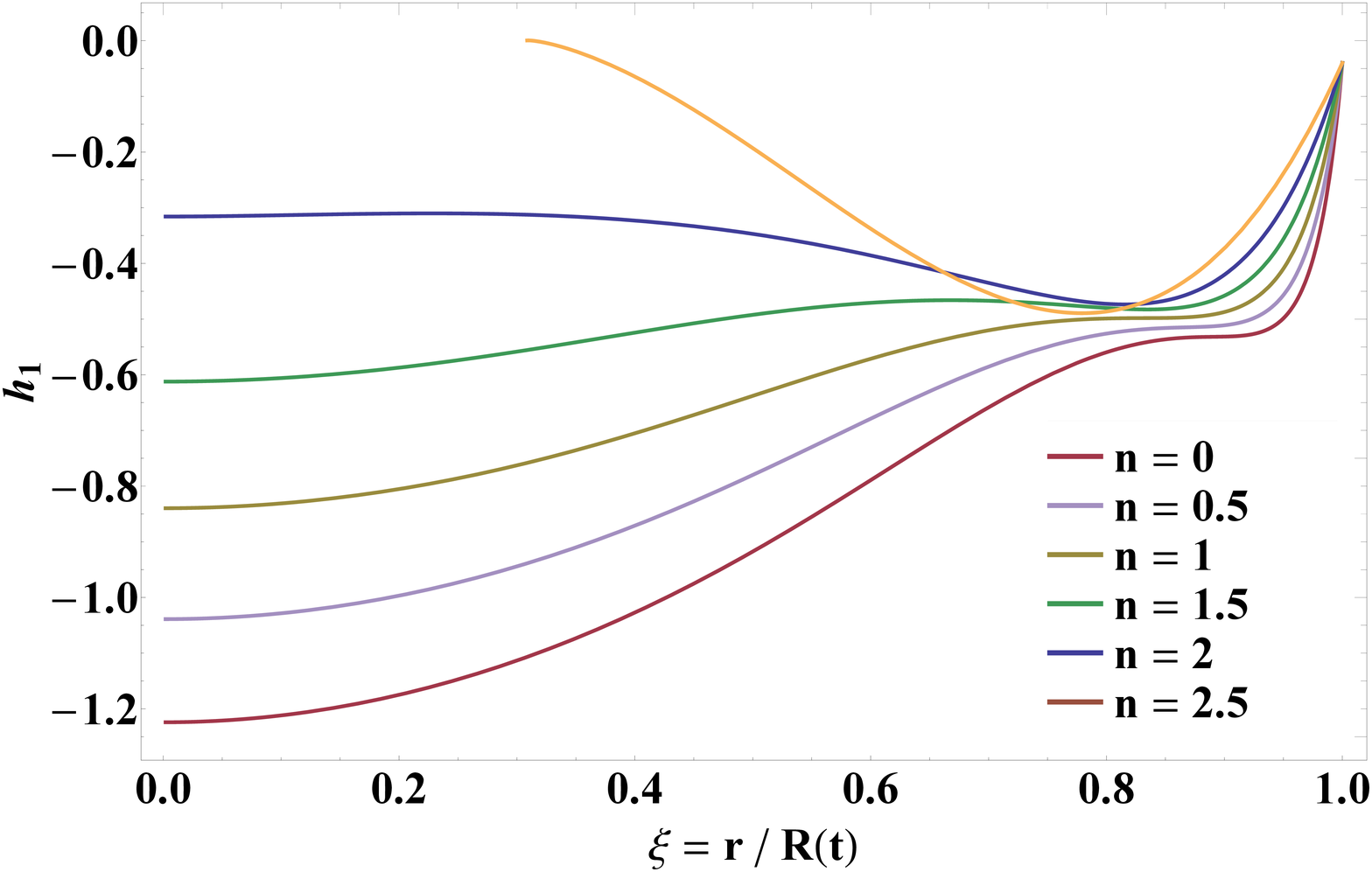} 
   \caption{Left: The relativistic, self-similar correction to the post-shock four-velocity for $\gamma = 4/3$ and the range of $n$ shown in the legend, where $n$ characterizes the power-law falloff of the ambient density with radius (i.e., $\rho'_{\rm a} \propto r^{-n}$). For larger values of $n$, the solution ends at a contact discontinuity at a finite $\xi_{\rm c}$, which is also where the Sedov-Taylor solution terminates. Right: The self-similar, relativistic correction to the pressure of the post-shock fluid. As the density profile steepens from a constant density to $\rho' \propto r^{-2}$, the pressure corrections become less severe, and the reduction of the post-shock pressure immediately behind the shock becomes less pronounced. When $n = 2.5$, the pressure equals zero at a contact discontinuity, and the magnitude of the correction shows a slight increase. }
   \label{fig:f1_g43_of_n}
\end{figure}

\begin{figure}[htbp] 
   \centering
   \includegraphics[width=0.495\textwidth]{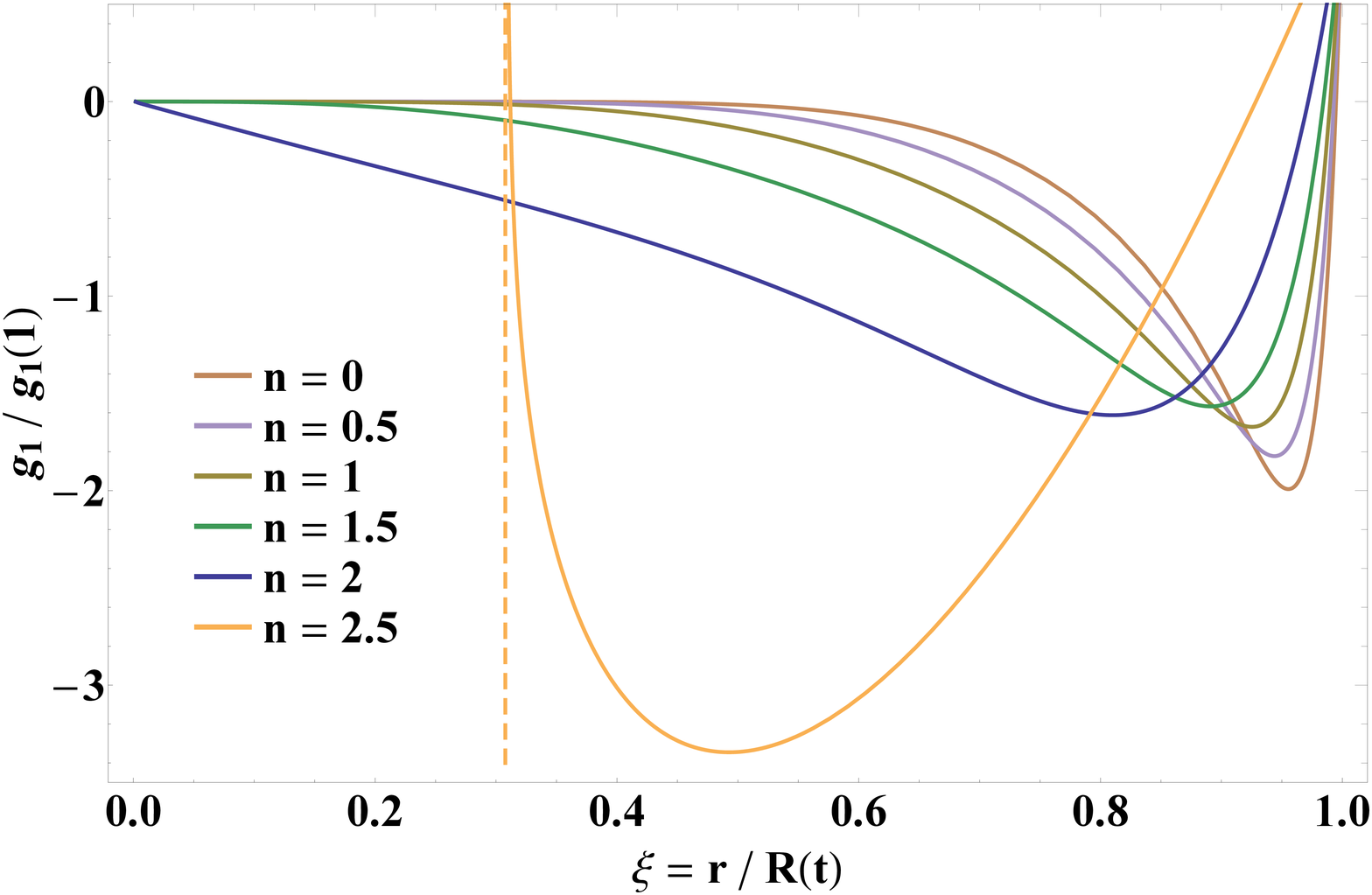} 
    \includegraphics[width=0.495\textwidth]{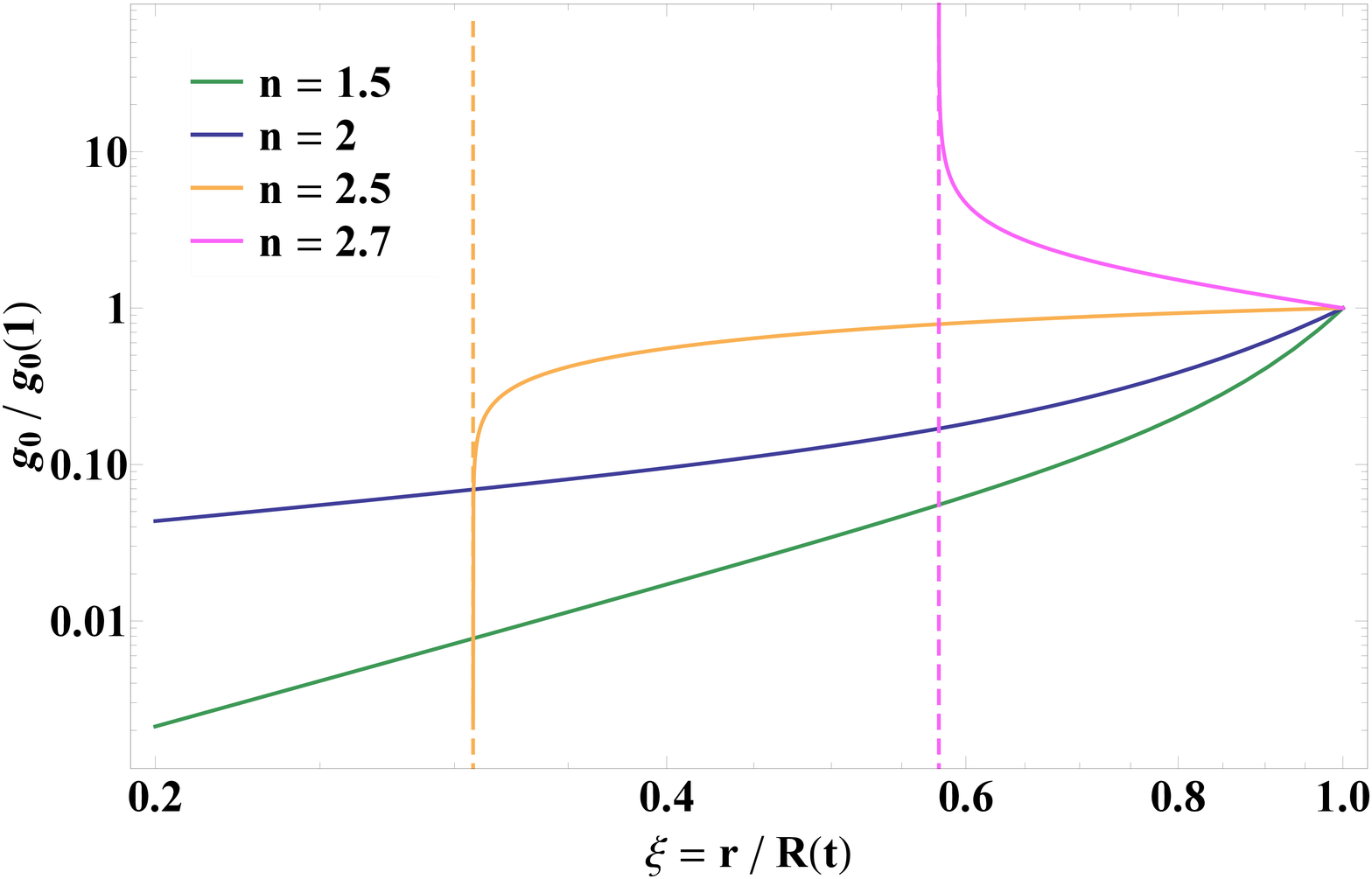} 
   \caption{Left: The relativistic, self-similar correction to the comoving density behind the shock for $\gamma = 4/3$ and the range of $n$ shown in the legend, where the density of the ambient medium $\rho'_{\rm a}$ falls off with spherical radius $r$ as $\rho'_{\rm a} \propto r^{-n}$. When the non-relativistic solution extends all the way to the origin, which occurs when $n\le 2$, the relativistic correction to the density approaches zero near $\xi = r = 0$. However, when the solution terminates at a contact discontinuity, which occurs for $n = 2.5$, the correction to the density diverges weakly at that point (which is at the location of the vertical, dashed line). Right: The non-relativistic, self-similar solution for the density behind the shock for the range of $n$ shown in the legend when the post-shock adiabatic index is $\gamma = 4/3$. The vertical, dashed lines indicate the locations of the contact discontinuity. When $n = 2.5$, the density at the contact discontinuity equals zero, while the Sedov-Taylor solution predicts a diverging density at the contact discontinuity for $n = 2.7$. Energy-conserving, relativistic corrections do not exist when the non-relativistic density either remains finite or diverges at the contact discontinuity. }
   \label{fig:g1_g43_of_n}
\end{figure}

\begin{figure}[htbp] 
   \centering
   \includegraphics[width=0.495\textwidth]{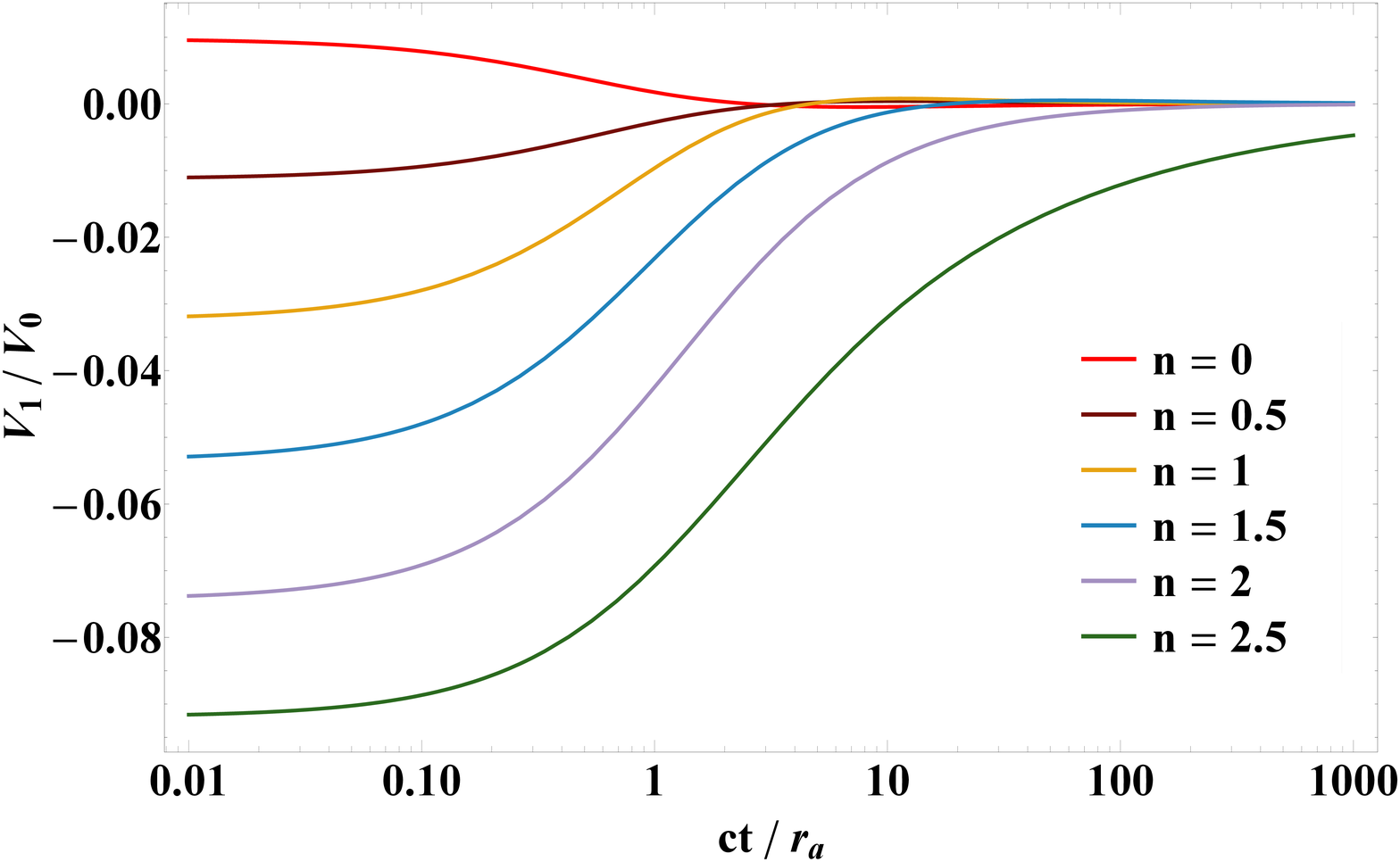} 
    \includegraphics[width=0.495\textwidth]{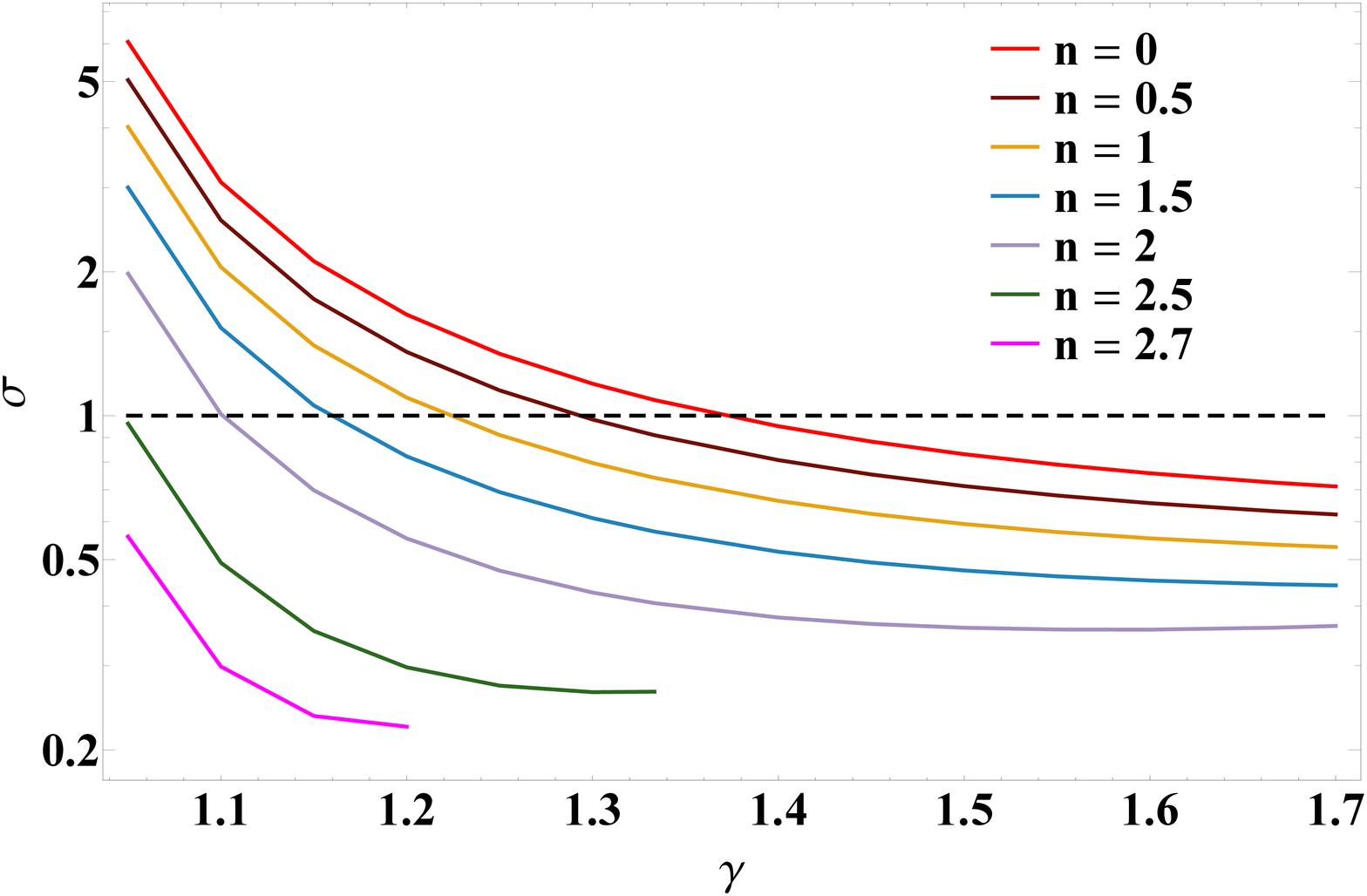} 
   \caption{Left: The relativistic correction to the shock three velocity $V_1$, normalized by the non-relativistic (Sedov-Taylor) solution $V_0$, as a function of time. Here we set $V_{\rm i}/c = 0.5$, where $V_{\rm i}$ is the velocity that the shock would have at $t = 0$ if relativistic effects were not included. Each curve corresponds to a different radial power-law index of the density of the ambient medium $n$, so the ambient density declines with radius as $\rho'_{\rm a} \propto r^{-n}$, as shown in the legend. As $n$ increases, relativistic effects are longer-lived owing to the fact that the non-relativistic, Sedov-Taylor shock speed falls off as a shallower function of time. Right: The eigenvalue $\sigma$, which relates the shock velocity to the explosion energy and ensures the conservation of that energy, as a function of the adiabatic index $\gamma$. Different curves are appropriate to the value of the power-law index of the density of the ambient medium given in the legend. Positive $\sigma$ implies that the shock speed is increased relative to the non-relativistic value in the comoving frame of the non-relativistic shock, i.e., observers moving with the Sedov-Taylor shock speed see an increase to the shock speed owing to relativistic effects. When $\sigma = 1$, time dilation and the relativistic boost to the shock speed in the comoving frame exactly balance to yield an observer-frame shock velocity that is identical to the Sedov-Taylor solution. For $\sigma < 1$, time dilation results in a three-velocity that is reduced compared to the Sedov-Taylor solution, and this can be seen directly from the left panel of this figure.}
   \label{fig:V1_g43_of_n}
\end{figure}

\begin{table}
\begin{center}
\begin{tabular}{|c|c|c|c|c|c|c|c|c|c|c|c|c|c|c|} \hline
\diagbox[width=1.5cm,height=1.5cm,outerrightsep=-8pt,dir = SE]{\hspace{-1.1cm}\vspace{.2cm} n}{$\gamma$} & 1.05 & 1.10 & 1.15 & 1.20 & 1.25 & 1.30 & 4/3 & 1.40 & 1.45 & 1.50 & 1.55 & 1.60 & 5/3 & 1.70\\
\hline
0 & $\sigma =$ 6.05 & 3.08 & 2.10 & 1.63 & 1.35 & 1.17 & {1.08} & {0.950} & 0.882 & 0.831 & 0.790 & 0.758 & 0.724 & 0.711 \\
\hline
0.5 & 5.04 & 2.56 & 1.75 & 1.36 & 1.13 & 0.982 & 0.910 & 0.807 & 0.753 & 0.712 & 0.681 & 0.656 & 0.631 & 0.621 \\
\hline
1 & 4.02 & 2.05 & 1.40 & 1.09 & 0.911 & 0.796 & 0.741 & 0.663 & 0.623 & 0.593 & 0.571 & 0.554 & 0.537 & 0.531 \\
\hline
1.5 & 3.00 & 1.53 & 1.05 & 0.822 & 0.692 & 0.610 & 0.572 & 0.519 & 0.493 & 0.474 & 0.461 & 0.452 & 0.444 & 0.442 \\
\hline
2 & 1.98 & 1.01 & 0.698 & 0.553 & 0.474 & 0.427 & 0.405 & 0.378 & 0.367 & 0.360 & 0.357 & 0.357 & 0.360 & 0.363 \\
\hline
2.5 & 0.962 & 0.492 & 0.354 & 0.298 & 0.272 & 0.264 & 0.265 & \ldots & \ldots & \ldots & \ldots & \ldots & \ldots & \ldots \\
\hline
2.7 & 0.559 & 0.299 & 0.235 & 0.224 & \ldots & \ldots & \ldots & \ldots & \ldots & \ldots & \ldots & \ldots & \ldots & \ldots \\
\hline
\end{tabular}
\caption{The eigenvalue $\sigma$ that ensures energy conservation as a function of $n$, which parameterizes the density profile of the ambient medium ($\rho' \propto r^{-n}$), and the post-shock adiabatic index of the gas $\gamma$. Values much greater than one indicate that relativistic effects are more important for less relativistic initial shock speeds, which are achieved for small $n$ and $\gamma$. Cells with an ellipsis correspond to instances where the Sedov-Taylor (non-relativistic) solution possesses a finite or diverging density at the contact discontinuity, for which we find no energy-conserving solution for the relativistic corrections.}
\label{tab:1}
\end{center}
\end{table}

Figure \ref{fig:f1_g43_of_n} illustrates the self-similar correction to the post-shock four-velocity for $\gamma = 4/3$ and the range of $n$ shown in the legend (the density profile of the ambient medium falls off with spherical radius $r$ as $\rho' \propto r^{-n}$). We see that, as $n$ increases, the relativistic correction to the velocity decreases in magnitude, and the solution with $n = 2.5$ ends at a contact discontinuity -- coinciding with a point in the flow where the non-relativistic velocity profile satisfies $f(\xi_{\rm c}) = \xi_{\rm c}$, such that the fluid elements at these locations are stationary with respect to the shock. The right panel of this figure shows the correction to the pressure profile behind the shock, again with $\gamma = 4/3$. As $n$ increases from $0$ to $2$, the pressure profile exhibits progressively less deviation behind the shock, and the overall magnitude of the correction is reduced. When $n = 2.5$, the post-shock correction to the pressure equals zero at the contact discontinuity, which is also where the non-relativistic pressure equals zero. 

The left panel of Figure \ref{fig:g1_g43_of_n} gives the correction to the self-similar, comoving density behind the shock for $\gamma = 4/3$ and the power-law indices of the ambient medium, $n$, in the legend. As for the pressure, the relativistic contributions tend to be less pronounced as $n$ increases from $0$ to 2, with less of the variation being confined to the immediate vicinity of the shock and the magnitude of the variation reduced. However, when $n = 2.5$, we see that the magnitude of the correction increases again, and the function $g_1$ actually diverges weakly at the location of the contact discontinuity. The right panel of this figure shows, for reference, the Sedov-Taylor solution for a subset of $n$, with the vertical, dashed lines coinciding with the location of the contact discontinuity. We see that, when $n = 2.5$, the non-relativistic density at the contact discontinuity equals zero, while for $n = 2.7$ the function $g_0$ diverges at the contact discontinuity. 

The relativistic correction to the lab-frame, three-velocity of the shock $V_1$, plotted relative to the non-relativistic velocity $V_0$ as a function of time, is shown in the left panel of Figure \ref{fig:V1_g43_of_n}; here we set $V_{\rm i} / c = 0.5$, where $V_{\rm i}$ is the Newtonian shock speed at $t = 0$. Different curves correspond to different radial power-law indices of the density of the ambient medium, $n$, such that $\rho'_{\rm a} \propto r^{-n}$. The correction is negative for larger values of $n$, which signifies that time dilation actually reduces the shock speed below the Sedov-Taylor prediction in spite of the fact that the velocity is increased in the comoving frame of the non-relativistic shock. Relativistic effects remain important for longer times as $n$ increases, and this occurs because the Sedov-Taylor velocity -- to the square of which, as shown in Equation \eqref{V1eq}, the relativistic correction is proportional -- declines less slowly as the density profile of the ambient medium steepens (which arises physically from the reduced momentum flux across the shock for larger $n$). 

The eigenvalue $\sigma$, which conserves the relativistic energy behind the shock and relates the shock speed to the explosion energy, is shown as a function of $\gamma$ in the right panel of Figure \ref{fig:V1_g43_of_n}. Each curve corresponds to the power-law index of the ambient medium, $n$, shown in the legend, and the horizontal, dashed line simply indicates where $\sigma = 1$ for clarity. The fact that $\sigma$ is always positive implies that observers moving with the non-relativistic, Sedov-Taylor shock speed see a relativistically-boosted shock velocity in that frame. However, it is only for $\sigma >1$ that observers in the lab frame (i.e., the frame in which the ambient medium is at rest) also measure a positive, relativistic increase to the shock three-velocity; for $\sigma < 1$ time dilation reduces the velocity below the Sedov-Taylor solution, as can be seen directly in the left panel of this figure. The values of $\sigma(n,\gamma)$ which were used to make this figure are given in Table \ref{tab:1}. 

It can be shown from the integrals of Equations \eqref{g0eq} -- \eqref{h0eq} that the self-similar, Sedov-Taylor density scales as 

\begin{equation}
g_0(\xi) \propto \left(1-f_0/\xi\right)^{\frac{6-n-n\gamma}{n+3\gamma-6}},
\end{equation}
which implies that when the Sedov-Taylor solution ends in a contact discontinuity, the density diverges at that location if $n > 6/(\gamma+1) \simeq 2.57$ for $\gamma = 4/3$ (see also \citealt{goodman90}). We do not find any solution for $\sigma$ that satisfies global energy conservation for power-law indices that are steeper than this value. Mathematically, solutions in this regime do not exist because we can combine the integral constraint on the energy \eqref{eigenint} and Equation \eqref{g1ex} to show that, if the relativistic correction to the energy is to remain finite, then we must have\footnote{Note that this is not required when the density equals zero at the contact discontinuity, as in this situation the density scales as $g_1 \propto (1-\xi_{\rm c})^{-\alpha}$ near $\xi_{\rm c}$ with $\alpha < 1$; thus, while the correction to the comoving density diverges at $\xi_{\rm c}$, it does so in a way that yields a finite, relativistic correction to the energy.} $f_1(\xi_{\rm c}) = \xi_{\rm c}(\xi_{\rm c}^2-1)/2$. However, from Equation \eqref{bc4} and the fact that $h_0(\xi_{\rm c}) = 0$, we see that we must also have $h_1(\xi_{\rm c}) = 0$ to maintain energy conservation. The system is therefore over-constrained when the density diverges at the contact discontinuity, and these two boundary conditions at $\xi_{\rm c}$ will not, in general, be satisfied simultaneously for a single $\sigma$. In particular, we find that the solution for which $f_1(\xi_{\rm c}) = \xi_{\rm c}(\xi_{\rm c}^2-1)$ possesses a finite, but non-zero pressure at the contact discontinuity, meaning that these solutions cannot simultaneously satisfy both the integral constraint \eqref{eigenint} and Equation \eqref{bc4}. 

It is also not surprising from a physical standpoint that these diverging-density solutions are problematic. For one, it was shown by \citet{goodman90} that such solutions are unstable to aspherical perturbations, as the decelerating nature of the fluid and the density inversion renders the contact discontinuity susceptible to the Rayleigh-Taylor instability. These Sedov-Taylor blastwaves therefore cannot be manifested in any physical (i.e., one with permissible angular deviations from spherical symmetry) scenario. The diverging-density solutions also violate the self-similar hypothesis that the flow is predominantly characterized by the physical conditions at a single point within the flow: for small $n$ and $\gamma$, the vast majority of the mass is contained very near the shock front. However, when the density profile becomes inverted, most of the mass is concentrated near the contact discontinuity, and the causal connectedness of the shocked fluid implies that the shock ``knows'' this property of the inner flow. It is therefore likely that, for the diverging-density solutions, the physical conditions both at the shock front and near the contact discontinuity remain important for establishing the long-term behavior of the shock, and the solution may be fundamentally non-self-similar.

It is interesting to note that the ultra-relativistic, Blandford-McKee blastwave avoids this issue, as the comoving pressure behind the shock transforms as a higher power of the shock Lorentz factor than the comoving density. The contribution of the kinetic energy to the total energy behind the blastwave is therefore dropped in their solution, and only the internal energy must be conserved. Thus, while the Blandford-McKee solution formally exists in this regime, it is likely that the \emph{integrated} kinetic energy remains important for the dynamics (i.e., by virtue of the boundary conditions at the shock, it is true in the ultra-relativistic limit that the kinetic energy is sub-dominant to the internal energy, but that may not be true deeper within the flow; the non-existence of relativistic corrections to the Sedov-Taylor solutions suggests that this may be the case for, at least, smaller values of the Lorentz factor).

Finally, one could argue that our prescription for the behavior of the relativistic corrections was too restrictive, and that while solutions of the form given by Equations \eqref{Uss} -- \eqref{pss} do not exist in this regime, there may be other, more general solutions that do satisfy energy conservation and the modified boundary conditions at the shock. Instead of writing the expressions for the perturbations as in \eqref{Uss} -- \eqref{pss}, we could have parameterized them as

\begin{equation}
U = U_{\rm s}\left\{f_0(\xi)+U_{\rm s}^2f_1(\xi,\chi)\right\},
\end{equation}
and similarly for the other variables, and therefore maintained the derivatives with respect to $\chi$ in Equations \eqref{cont1} -- \eqref{rmom1} ({recall that $\chi = \ln R$ is a time-like variable)}. However, it is difficult to see how relativistic effects (to order $V^2/c^2$) could modify the solution in a way \emph{other} than one that scales as $V^2/c^2$, as this is the only physical smallness parameter introduced in the problem. Therefore, any additional time dependence in the expression $f_1(\xi,\chi)$ should, instead, be regarded as a higher-order correction to the self-similar solution; we also note that this is precisely the motivation for expanding the self-similar functions as $f(\xi,\chi) \simeq f_0(\xi)+U_{\rm s}^2f_1(\xi)$ -- the second term accounts for the time dependence that is induced by relativistic effects, and that time dependence can only physically be of the form $U_{\rm s}^2$. While we acknowledge that this is not a rigorous proof of the non-existence of more general solutions, we find it suggestive that no such solutions exist. 

\section{Summary and Conclusions}
\label{sec:summary}
When the shockwave from an astrophysical explosion is strong (Mach number much greater than one) and non-relativistic, and therefore characterized by a shock speed $V$ much less than the speed of light $c$, the Sedov-Taylor, energy-conserving blastwave provides an analytic, self-similar solution for the temporal evolution of the shock itself, and the time and space-dependent evolution of the post-shock velocity, density, and pressure. In the other, extreme-relativistic limit where $V \simeq c$, the Blandford-McKee blastwave gives the energy-conserving evolution of the shock Lorentz factor and the post-shock fluid quantities. In between these two extremes, the finite speed of light introduces an additional velocity scale into the problem, which destroys the pure self-similarity of the solutions.

In this paper, we analyzed the leading-order, relativistic corrections to the fluid equations -- which enter as $\mathcal{O}(V^2/c^2)$ -- to understand the effects that such relativistic terms have on the non-relativistic, Sedov-Taylor solution for strong shock propagation. By treating such terms as perturbations (i.e., ignoring nonlinear terms that enter as higher powers of $V^2/c^2$), we showed that there are relativistic corrections to the Sedov-Taylor solutions for the post-shock fluid quantities that vary (consistent with expectations) as $V^2/c^2$. In particular, we demonstrated that the radial component of the post-shock, fluid four-velocity can be written as $U = U_{\rm s}\left\{f_0(\xi)+U_{\rm s}^2/c^2\times f_1(\xi)\right\}$, where $U_{\rm s}$ is the shock four-velocity, $f_0$ is the Sedov-Taylor solution, $\xi = r/R(t)$ with $R(t)$ the shock position, and $f_1$ is a function that is self-consistently determined from the fluid equations and the relativistic jump conditions at the shock. The function $f_1$, and the analogous functions $g_1$ and $h_1$ for the density and pressure, respectively, induce more nonlinear behavior to the post-shock velocity, further compress the post-shock material to the immediate vicinity of the shock itself, and generate greater variation in the post-shock pressure as compared to the Sedov-Taylor, non-relativistic limit (see Figures \ref{fig:f0g0h0_n0_g43} -- \ref{fig:lab_rho_n0_g43}). These are all features of the ultra-relativistic, Blandford-McKee blastwave, where the pressure declines rapidly behind the shock and all of the material is swept into a shell of width $\Delta R = R/U_{\rm s}^2$.

In addition to the post-shock fluid quantities, we also determined the relativistic correction to the velocity of the shock itself. We denoted this additional correction by an ``eigenvalue'' $\sigma$, such that the relativistically-corrected shock four-velocity is written implicitly as $U_{\rm s}^2R^{3-n} = E\left(1+\sigma U_{\rm s}^2/c^2\right)$. When $\sigma \equiv 0$, one recovers the familiar relationship between the shock velocity and position that guarantees the conservation of the blast energy, $E$. However, owing to the existence of relativistic corrections to the energy, $\sigma$ cannot be exactly zero, and there must therefore be corrections to the shock velocity that maintain total (i.e., including relativistic terms to order $U_{\rm s}^2/c^2$) energy conservation. For all of our solutions, the value of $\sigma$ was found to be positive, implying an increase to the shock four-velocity from relativistic effects; equivalently, observers moving at the non-relativistic shock speed measure a small, slight increase to the shock velocity, and hence the true shock position leads the non-relativistic one in the comoving frame of the non-relativistic shock. Nevertheless, the lab-frame three-velocity is reduced from the four-velocity by time dilation, which counterbalances the relativistically-boosted effect of positive $\sigma$, and if $\sigma \equiv 1$ these two effects exactly offset to yield a shock three-velocity that is identical to the Newtonian, Sedov-Taylor value. Interestingly, for a radiation-pressure dominated, $\gamma = 4/3$ post-shock equation of state and a constant-density ambient medium, the shock three-velocity is slightly increased over the non-relativistic value, while for all declining density profiles with $n > 0.5$ -- where the ambient density $\rho'_{\rm a}$ falls off with spherical radius $r$ as $\rho'_{\rm a} \propto r^{-n}$ -- the lab-frame velocity is reduced below the non-relativistic value (see Figure \ref{fig:V1_g43_of_n} and Table \ref{tab:1} for the values of $\sigma$ for a range of $n$ and $\gamma$). 

Once the power-law index of the ambient density profile equals or exceeds the critical value $n_{\rm cr}(\gamma) = 6/(\gamma+1)$, the density at the contact discontinuity present in the Sedov-Taylor solution diverges. For Sedov-Taylor blastwaves with $n > n_{\rm cr}$, we do not find any solution for $\sigma$ that maintains a finite and conserved relativistic correction to the energy, which mathematically follows directly from the nature of the solutions for $f_1$, $g_1$, and $h_1$ and the integral that maintains the conservation of energy (see Equations \ref{eigenint} and \ref{bc4}). It is possible that more general, time-dependent solutions (i.e., those that do not assume the form given by Equations \ref{Uss} -- \ref{pss}) could be found in this regime that do conserve energy. However, based on the argument that additional time dependence from the self-similar, Sedov-Taylor solutions is itself seeded by relativistic effects, it is difficult to see how any extra time dependence would not be in the form of higher-order (than $U_{\rm s}^2/c^2$) terms. We therefore find it unlikely that such generalized solutions exist at the leading relativistic order.

Here we described the leading-order, relativistic corrections to the fluid flow, which enter into the fluid equations and the boundary conditions as $U_{\rm s}^2/c^2$. One can regard our solutions as the first in a series expansion of the fluid equations in $U_{\rm s}^2$, and the next-order solution for (for example) the four-velocity would be $U = U_{\rm s}\left\{f_0(\xi)+U_{\rm s}^2f_1(\xi)+U_{\rm s}^4f_2(\xi)\right\}$; one could then, by expanding the fluid equations to the next order, derive self-consistent equations for $f_2$, $g_2$, and $h_2$, and the boundary conditions at the shock could be found by expanding the general jump conditions to the next-highest order. In principle, one should also be able to construct the next-order solution for the Blandford-McKee solution, and approach the problem from the other, ultrarelativistic direction by finding the next-highest-order correction in $1/\Gamma$. 

In this paper we focused on ambient density profiles less steep than $\rho'_{\rm a} \propto r^{-3}$. For steeper density profiles, the shock enters an accelerating regime, and the self-similar solutions for the post-shock fluid quantities are provided by \citet{waxman93} (see also \citealt{koo90}). In this case, the self-similar flow is constrained to lie between a sonic point within the interior of the flow and the shock front, and energy and mass are drained into a non-self-similar, inner region that is causally disconnected from the fluid near the shock. One could apply all of the formalism developed in this paper to derive the leading-order, relativistic corrections to such accelerating, self-similar flows, and the resulting equations for the functions $f_1$, $g_1$, and $h_1$ would, in fact, appear almost identical to Equations \eqref{cont1} -- \eqref{rmom1}, with the exception that various factors of $n-3$ -- which result from energy conservation -- would be replaced by a numerical factor that is determined from the self-similar, accelerating solutions. When the shock accelerates, such corrections actually come to \emph{dominate} the self-similar solution at late times, owing to the increasing nature of $U_{\rm s}^2/c^2$, and one could interpret this result by saying that such self-similar solutions are ``unstable'' to relativistic corrections.

These solutions for the relativistically-corrected shock speed and post-shock fluid quantities could be used for generating more accurate models for the late-time lightcurves of long gamma-ray bursts as well as the lightcurves of energetic supernovae. In particular, the relativistic beaming induced by the marginally-relativistic velocity, and the prediction for the time over which the shock speed declines to sub-relativistic speeds, would yield correspondingly different break timescales for the lightcurve of the event and peaks in the synchrotron spectrum (e.g., \citealt{sari98, decolle12}). The event AT2018cow \citep{prentice18, rivera18, ho19, kuin19, margutti19, perley19}, tentatively an extreme example of the class of fast-rising transients \citep{drout14}, also provided evidence of a moderately relativistic outflow with speed $\sim 0.1c$; the model presented here could be combined with current multiwavelength data to further constrain properties of the progenitor and surrounding medium.

\acknowledgements
This work was supported by NASA through the Einstein Fellowship Program, Grant PF6-170170. I thank Brian Metzger, Eliot Quataert, and Jonathan Zrake for useful discussions.

\bibliographystyle{aasjournal}
\bibliography{refs}

\begin{thebibliography}{}
\expandafter\ifx\csname natexlab\endcsname\relax\def\natexlab#1{#1}\fi
\providecommand{\url}[1]{\href{#1}{#1}}
\providecommand{\dodoi}[1]{doi:~\href{http://doi.org/#1}{\nolinkurl{#1}}}
\providecommand{\doeprint}[1]{\href{http://ascl.net/#1}{\nolinkurl{http://ascl.net/#1}}}
\providecommand{\doarXiv}[1]{\href{https://arxiv.org/abs/#1}{\nolinkurl{https://arxiv.org/abs/#1}}}

\bibitem[{{Abbott} {et~al.}(2017){Abbott}, {Abbott}, {Abbott}, {Acernese},
  {Ackley}, {Adams}, {Adams}, {Addesso}, {Adhikari}, {Adya}, \&
  et~al.}]{abbott17c}
{Abbott}, B.~P., {Abbott}, R., {Abbott}, T.~D., {et~al.} 2017, \apjl, 848, L12,
  \dodoi{10.3847/2041-8213/aa91c9}

\bibitem[{{Arnett}(1982)}]{arnett82}
{Arnett}, W.~D. 1982, \apjl, 263, L55, \dodoi{10.1086/183923}

\bibitem[{{Bethe} \& {Wilson}(1985)}]{bethe85}
{Bethe}, H.~A., \& {Wilson}, J.~R. 1985, \apj, 295, 14, \dodoi{10.1086/163343}

\bibitem[{{Blandford} \& {McKee}(1976)}]{blandford76}
{Blandford}, R.~D., \& {McKee}, C.~F. 1976, Physics of Fluids, 19, 1130,
  \dodoi{10.1063/1.861619}

\bibitem[{{Blondin} {et~al.}(2003){Blondin}, {Mezzacappa}, \&
  {DeMarino}}]{blondin03}
{Blondin}, J.~M., {Mezzacappa}, A., \& {DeMarino}, C. 2003, \apj, 584, 971,
  \dodoi{10.1086/345812}

\bibitem[{{Burrows} {et~al.}(1995){Burrows}, {Hayes}, \& {Fryxell}}]{burrows95}
{Burrows}, A., {Hayes}, J., \& {Fryxell}, B.~A. 1995, \apj, 450, 830,
  \dodoi{10.1086/176188}

\bibitem[{{Chevalier}(1976)}]{chevalier76}
{Chevalier}, R.~A. 1976, \apj, 207, 872, \dodoi{10.1086/154557}

\bibitem[{{Colgate} \& {White}(1966)}]{colgate66}
{Colgate}, S.~A., \& {White}, R.~H. 1966, \apj, 143, 626,
  \dodoi{10.1086/148549}

\bibitem[{{Corsi} {et~al.}(2014){Corsi}, {Ofek}, {Gal-Yam}, {Frail},
  {Kulkarni}, {Fox}, {Kasliwal}, {Sullivan}, {Horesh}, {Carpenter}, {Maguire},
  {Arcavi}, {Cenko}, {Cao}, {Mooley}, {Pan}, {Sesar}, {Sternberg}, {Xu},
  {Bersier}, {James}, {Bloom}, \& {Nugent}}]{corsi14}
{Corsi}, A., {Ofek}, E.~O., {Gal-Yam}, A., {et~al.} 2014, \apj, 782, 42,
  \dodoi{10.1088/0004-637X/782/1/42}

\bibitem[{{Corsi} {et~al.}(2016){Corsi}, {Gal-Yam}, {Kulkarni}, {Frail},
  {Mazzali}, {Cenko}, {Kasliwal}, {Cao}, {Horesh}, {Palliyaguru}, {Perley},
  {Laher}, {Taddia}, {Leloudas}, {Maguire}, {Nugent}, {Sollerman}, \&
  {Sullivan}}]{corsi16}
{Corsi}, A., {Gal-Yam}, A., {Kulkarni}, S.~R., {et~al.} 2016, \apj, 830, 42,
  \dodoi{10.3847/0004-637X/830/1/42}

\bibitem[{{Coughlin} {et~al.}(2018{\natexlab{a}}){Coughlin}, {Quataert},
  {Fern{\'a}ndez}, \& {Kasen}}]{coughlin18a}
{Coughlin}, E.~R., {Quataert}, E., {Fern{\'a}ndez}, R., \& {Kasen}, D.
  2018{\natexlab{a}}, \mnras, 477, 1225, \dodoi{10.1093/mnras/sty667}

\bibitem[{{Coughlin} {et~al.}(2018{\natexlab{b}}){Coughlin}, {Quataert}, \&
  {Ro}}]{coughlin18b}
{Coughlin}, E.~R., {Quataert}, E., \& {Ro}, S. 2018{\natexlab{b}}, \apj, 863,
  158, \dodoi{10.3847/1538-4357/aad198}

\bibitem[{{Coughlin} {et~al.}(2019){Coughlin}, {Ro}, \&
  {Quataert}}]{coughlin19}
{Coughlin}, E.~R., {Ro}, S., \& {Quataert}, E. 2019, \apj, 874, 58,
  \dodoi{10.3847/1538-4357/ab09ec}

\bibitem[{{De Colle} {et~al.}(2012){De Colle}, {Ramirez-Ruiz}, {Granot}, \&
  {Lopez-Camara}}]{decolle12}
{De Colle}, F., {Ramirez-Ruiz}, E., {Granot}, J., \& {Lopez-Camara}, D. 2012,
  \apj, 751, 57, \dodoi{10.1088/0004-637X/751/1/57}

\bibitem[{{Drout} {et~al.}(2011){Drout}, {Soderberg}, {Gal-Yam}, {Cenko},
  {Fox}, {Leonard}, {Sand}, {Moon}, {Arcavi}, \& {Green}}]{drout11}
{Drout}, M.~R., {Soderberg}, A.~M., {Gal-Yam}, A., {et~al.} 2011, \apj, 741,
  97, \dodoi{10.1088/0004-637X/741/2/97}

\bibitem[{{Drout} {et~al.}(2014){Drout}, {Chornock}, {Soderberg}, {Sanders},
  {McKinnon}, {Rest}, {Foley}, {Milisavljevic}, {Margutti}, {Berger},
  {Calkins}, {Fong}, {Gezari}, {Huber}, {Kankare}, {Kirshner}, {Leibler},
  {Lunnan}, {Mattila}, {Marion}, {Narayan}, {Riess}, {Roth}, {Scolnic},
  {Smartt}, {Tonry}, {Burgett}, {Chambers}, {Hodapp}, {Jedicke}, {Kaiser},
  {Magnier}, {Metcalfe}, {Morgan}, {Price}, \& {Waters}}]{drout14}
{Drout}, M.~R., {Chornock}, R., {Soderberg}, A.~M., {et~al.} 2014, \apj, 794,
  23, \dodoi{10.1088/0004-637X/794/1/23}

\bibitem[{{Duffell} \& {MacFadyen}(2013)}]{duffell13}
{Duffell}, P.~C., \& {MacFadyen}, A.~I. 2013, \apj, 775, 87,
  \dodoi{10.1088/0004-637X/775/2/87}

\bibitem[{{Fern{\'a}ndez} {et~al.}(2018){Fern{\'a}ndez}, {Quataert},
  {Kashiyama}, \& {Coughlin}}]{fernandez18}
{Fern{\'a}ndez}, R., {Quataert}, E., {Kashiyama}, K., \& {Coughlin}, E.~R.
  2018, \mnras, 476, 2366, \dodoi{10.1093/mnras/sty306}

\bibitem[{{Goodman}(1990)}]{goodman90}
{Goodman}, J. 1990, \apj, 358, 214, \dodoi{10.1086/168977}

\bibitem[{{Ho} {et~al.}(2019){Ho}, {Phinney}, {Ravi}, {Kulkarni}, {Petitpas},
  {Emonts}, {Bhalerao}, {Blundell}, {Cenko}, {Dobie}, {Howie}, {Kamraj},
  {Kasliwal}, {Murphy}, {Perley}, {Sridharan}, \& {Yoon}}]{ho19}
{Ho}, A.~Y.~Q., {Phinney}, E.~S., {Ravi}, V., {et~al.} 2019, \apj, 871, 73,
  \dodoi{10.3847/1538-4357/aaf473}

\bibitem[{{Kobayashi} \& {Sari}(2000)}]{kobayashi00}
{Kobayashi}, S., \& {Sari}, R. 2000, \apj, 542, 819, \dodoi{10.1086/317021}

\bibitem[{{Koo} \& {McKee}(1990)}]{koo90}
{Koo}, B.-C., \& {McKee}, C.~F. 1990, \apj, 354, 513, \dodoi{10.1086/168712}

\bibitem[{{Kuin} {et~al.}(2019){Kuin}, {Wu}, {Oates}, {Lien}, {Emery},
  {Kennea}, {de Pasquale}, {Han}, {Brown}, {Tohuvavohu}, {Breeveld}, {Burrows},
  {Cenko}, {Campana}, {Levan}, {Markwardt}, {Osborne}, {Page}, {Page},
  {Sbarufatti}, {Siegel}, \& {Troja}}]{kuin19}
{Kuin}, N.~P.~M., {Wu}, K., {Oates}, S., {et~al.} 2019, \mnras,
  \dodoi{10.1093/mnras/stz053}

\bibitem[{{Levinson} {et~al.}(2002){Levinson}, {Ofek}, {Waxman}, \&
  {Gal-Yam}}]{levinson02}
{Levinson}, A., {Ofek}, E.~O., {Waxman}, E., \& {Gal-Yam}, A. 2002, \apj, 576,
  923, \dodoi{10.1086/341866}

\bibitem[{{Li} \& {Paczy{\'n}ski}(1998)}]{li98}
{Li}, L.-X., \& {Paczy{\'n}ski}, B. 1998, \apjl, 507, L59,
  \dodoi{10.1086/311680}

\bibitem[{{Lovegrove} \& {Woosley}(2013)}]{lovegrove13}
{Lovegrove}, E., \& {Woosley}, S.~E. 2013, \apj, 769, 109,
  \dodoi{10.1088/0004-637X/769/2/109}

\bibitem[{{MacFadyen} \& {Woosley}(1999)}]{macfadyen99}
{MacFadyen}, A.~I., \& {Woosley}, S.~E. 1999, \apj, 524, 262,
  \dodoi{10.1086/307790}

\bibitem[{{Margutti} {et~al.}(2019){Margutti}, {Metzger}, {Chornock}, {Vurm},
  {Roth}, {Grefenstette}, {Savchenko}, {Cartier}, {Steiner}, {Terreran},
  {Margalit}, {Migliori}, {Milisavljevic}, {Alexander}, {Bietenholz},
  {Blanchard}, {Bozzo}, {Brethauer}, {Chilingarian}, {Coppejans}, {Ducci},
  {Ferrigno}, {Fong}, {G{\"o}tz}, {Guidorzi}, {Hajela}, {Hurley}, {Kuulkers},
  {Laurent}, {Mereghetti}, {Nicholl}, {Patnaude}, {Ubertini}, {Banovetz},
  {Bartel}, {Berger}, {Coughlin}, {Eftekhari}, {Frederiks}, {Kozlova},
  {Laskar}, {Svinkin}, {Drout}, {MacFadyen}, \& {Paterson}}]{margutti19}
{Margutti}, R., {Metzger}, B.~D., {Chornock}, R., {et~al.} 2019, \apj, 872, 18,
  \dodoi{10.3847/1538-4357/aafa01}

\bibitem[{{M{\"o}sta} {et~al.}(2015){M{\"o}sta}, {Ott}, {Radice}, {Roberts},
  {Schnetter}, \& {Haas}}]{mosta15}
{M{\"o}sta}, P., {Ott}, C.~D., {Radice}, D., {et~al.} 2015, \nat, 528, 376,
  \dodoi{10.1038/nature15755}

\bibitem[{{Nadezhin}(1980)}]{nadezhin80}
{Nadezhin}, D.~K. 1980, \apss, 69, 115, \dodoi{10.1007/BF00638971}

\bibitem[{{Nakar} \& {Piran}(2011)}]{nakar11}
{Nakar}, E., \& {Piran}, T. 2011, \nat, 478, 82, \dodoi{10.1038/nature10365}

\bibitem[{{Ostriker} \& {McKee}(1988)}]{ostriker88}
{Ostriker}, J.~P., \& {McKee}, C.~F. 1988, Reviews of Modern Physics, 60, 1,
  \dodoi{10.1103/RevModPhys.60.1}

\bibitem[{{Perley} {et~al.}(2019){Perley}, {Mazzali}, {Yan}, {Cenko}, {Gezari},
  {Taggart}, {Blagorodnova}, {Fremling}, {Mockler}, {Singh}, {Tominaga},
  {Tanaka}, {Watson}, {Ahumada}, {Anupama}, {Ashall}, {Becerra}, {Bersier},
  {Bhalerao}, {Bloom}, {Butler}, {Copperwheat}, {Coughlin}, {De}, {Drake},
  {Duev}, {Frederick}, {Gonz{\'a}lez}, {Goobar}, {Heida}, {Ho}, {Horst},
  {Hung}, {Itoh}, {Jencson}, {Kasliwal}, {Kawai}, {Khanam}, {Kulkarni},
  {Kumar}, {Kumar}, {Kutyrev}, {Lee}, {Maeda}, {Mahabal}, {Murata}, {Neill},
  {Ngeow}, {Penprase}, {Pian}, {Quimby}, {Ramirez-Ruiz}, {Richer},
  {Rom{\'a}n-Z{\'u}{\~n}iga}, {Sahu}, {Srivastav}, {Socia}, {Sollerman},
  {Tachibana}, {Taddia}, {Tinyanont}, {Troja}, {Ward}, {Wee}, \&
  {Yu}}]{perley19}
{Perley}, D.~A., {Mazzali}, P.~A., {Yan}, L., {et~al.} 2019, \mnras, 484, 1031,
  \dodoi{10.1093/mnras/sty3420}

\bibitem[{{Piro}(2013)}]{piro13}
{Piro}, A.~L. 2013, \apjl, 768, L14, \dodoi{10.1088/2041-8205/768/1/L14}

\bibitem[{{Prentice} {et~al.}(2018){Prentice}, {Maguire}, {Smartt}, {Magee},
  {Schady}, {Sim}, {Chen}, {Clark}, {Colin}, {Fulton}, {McBrien}, {O'Neill},
  {Smith}, {Ashall}, {Chambers}, {Denneau}, {Flewelling}, {Heinze}, {Holoien},
  {Huber}, {Kochanek}, {Mazzali}, {Prieto}, {Rest}, {Shappee}, {Stalder},
  {Stanek}, {Stritzinger}, {Thompson}, \& {Tonry}}]{prentice18}
{Prentice}, S.~J., {Maguire}, K., {Smartt}, S.~J., {et~al.} 2018, \apjl, 865,
  L3, \dodoi{10.3847/2041-8213/aadd90}

\bibitem[{{Rivera Sandoval} {et~al.}(2018){Rivera Sandoval}, {Maccarone},
  {Corsi}, {Brown}, {Pooley}, \& {Wheeler}}]{rivera18}
{Rivera Sandoval}, L.~E., {Maccarone}, T.~J., {Corsi}, A., {et~al.} 2018,
  \mnras, 480, L146, \dodoi{10.1093/mnrasl/sly145}

\bibitem[{{Sari} {et~al.}(1998){Sari}, {Piran}, \& {Narayan}}]{sari98}
{Sari}, R., {Piran}, T., \& {Narayan}, R. 1998, \apjl, 497, L17,
  \dodoi{10.1086/311269}

\bibitem[{{Sedov}(1959)}]{sedov59}
{Sedov}, L.~I. 1959, {Similarity and Dimensional Methods in Mechanics}

\bibitem[{{Soderberg} {et~al.}(2006){Soderberg}, {Kulkarni}, {Nakar}, {Berger},
  {Cameron}, {Fox}, {Frail}, {Gal-Yam}, {Sari}, {Cenko}, {Kasliwal},
  {Chevalier}, {Piran}, {Price}, {Schmidt}, {Pooley}, {Moon}, {Penprase},
  {Ofek}, {Rau}, {Gehrels}, {Nousek}, {Burrows}, {Persson}, \&
  {McCarthy}}]{soderberg06}
{Soderberg}, A.~M., {Kulkarni}, S.~R., {Nakar}, E., {et~al.} 2006, \nat, 442,
  1014, \dodoi{10.1038/nature05087}

\bibitem[{{Taylor}(1950)}]{taylor50}
{Taylor}, G. 1950, Proceedings of the Royal Society of London Series A, 201,
  159, \dodoi{10.1098/rspa.1950.0049}

\bibitem[{{Waxman} \& {Shvarts}(1993)}]{waxman93}
{Waxman}, E., \& {Shvarts}, D. 1993, Physics of Fluids A, 5, 1035,
  \dodoi{10.1063/1.858668}

\bibitem[{{Whitesides} {et~al.}(2017){Whitesides}, {Lunnan}, {Kasliwal},
  {Perley}, {Corsi}, {Cenko}, {Blagorodnova}, {Cao}, {Cook}, {Doran},
  {Frederiks}, {Fremling}, {Hurley}, {Karamehmetoglu}, {Kulkarni}, {Leloudas},
  {Masci}, {Nugent}, {Ritter}, {Rubin}, {Savchenko}, {Sollerman}, {Svinkin},
  {Taddia}, {Vreeswijk}, \& {Wozniak}}]{whitesides17}
{Whitesides}, L., {Lunnan}, R., {Kasliwal}, M.~M., {et~al.} 2017, \apj, 851,
  107, \dodoi{10.3847/1538-4357/aa99de}

\bibitem[{{Woosley}(1993)}]{woosley93}
{Woosley}, S.~E. 1993, \apj, 405, 273, \dodoi{10.1086/172359}

\bibitem[{{Woosley} \& {Bloom}(2006)}]{woosley06}
{Woosley}, S.~E., \& {Bloom}, J.~S. 2006, \araa, 44, 507,
  \dodoi{10.1146/annurev.astro.43.072103.150558}

\bibitem[{{Xie} {et~al.}(2018){Xie}, {Zrake}, \& {MacFadyen}}]{xie18}
{Xie}, X., {Zrake}, J., \& {MacFadyen}, A. 2018, \apj, 863, 58,
  \dodoi{10.3847/1538-4357/aacf9c}

\end{thebibliography}

\end{document}